\documentclass[%
reprint,
superscriptaddress,
longbibliography,
 amsmath,amssymb,
aps,
prstab,
]{revtex4-1}

\usepackage{graphicx}
\usepackage{dcolumn}
\usepackage{bm}
\usepackage{lineno,hyperref,amsfonts,amssymb}
\usepackage{epstopdf}
\usepackage{afterpage}

\begin{document}

\title{Correction of magnetic optics and beam trajectory using LOCO based algorithm with expanded experimental data sets}
\thanks{Operated by Fermi Research Alliance, LLC under Contract No. DE-AC02-07CH11359 with the United States Department of Energy.}

\affiliation{Fermi National Accelerator Laboratory, Batavia, Illinois 60510, USA}
\affiliation{Budker Institute for Nuclear Physics, Novosibirsk, 630090, Russia}

\author{A.~Romanov} \affiliation{Fermi National Accelerator Laboratory, Batavia, Illinois 60510, USA}\affiliation{Budker Institute for Nuclear Physics, Novosibirsk, 630090, Russia}
\author{D.~Edstrom~Jr.} \affiliation{Fermi National Accelerator Laboratory, Batavia, Illinois 60510, USA}
\author{F.~A.~Emanov} \affiliation{Budker Institute for Nuclear Physics, Novosibirsk, 630090, Russia}
\author{I.~A.~Koop} \affiliation{Budker Institute for Nuclear Physics, Novosibirsk, 630090, Russia}
\author{E.~A.~Perevedentsev} \affiliation{Budker Institute for Nuclear Physics, Novosibirsk, 630090, Russia}
\author{Yu.~A.~Rogovsky} \affiliation{Budker Institute for Nuclear Physics, Novosibirsk, 630090, Russia}
\author{D.~B.~Shwartz} \affiliation{Budker Institute for Nuclear Physics, Novosibirsk, 630090, Russia}
\author{A.~Valishev} \affiliation{Fermi National Accelerator Laboratory, Batavia, Illinois 60510, USA}

\date{\today}

\begin{abstract}
Precise beam based measurement and correction of magnetic optics is essential for the successful operation of accelerators. The LOCO algorithm is a proven and reliable tool, which in some situations can be improved by using a broader class of experimental data. The standard data sets for LOCO include the closed orbit responses to dipole corrector variation, dispersion, and betatron tunes. This paper discusses the benefits from augmenting the data with four additional classes of experimental data: the beam shape measured with beam profile monitors; responses of closed orbit bumps to focusing field variations; betatron tune responses to focusing field variations; BPM-to-BPM betatron phase advances and beta functions in BPMs from turn-by-turn coordinates of kicked beam. All of the described features were implemented in the Sixdsimulation software that was used to correct the optics of the VEPP-2000 collider, the VEPP-5 injector booster ring, and the FAST linac.

\end{abstract}

\maketitle


\section{Introduction}
Successful operation of modern particle accelerators requires precise control over the magnetic fields along the path of charged particles. Strong focusing schemes, high element density, and many other aspects make the task of building a machine to specified requirements in terms of beam properties on the first beam run almost impossible and certainly very expensive. With sufficient beam instrumentation and correction capabilities in the accelerator, it is possible to partially substitute the tedious direct characterization of magnetic fields with beam based methods.

Beam based methods can be divided into two groups. The first group of methods can provide information about beam without knowing the detailed model of the accelerator. The second group is aimed at obtaining a complete model of the machine. Methods from the first group differ by complexity, from measuring betatron tunes, to obtaining beta-functions from analysis of kicked beam oscillation and many others. The LOCO method is a well tested and reliable example from the second group \cite{Safranek:1997mra,Sajaev:2005qw,Resende:2010cfa,Liu:2010ula,Benedetti:2011za,Roblin:2011zz,Aiba:2011zb,Martin:2014zpa,Smaluk:2016vgy,Ji:2016ayy}. It is based on the analysis of closed orbit responses to dipole corrector variations. The model of accelerator is adjusted so that it best describes the measured data. The robustness of LOCO is based on the significant redundancy of experimental data.

One of the key requirements of LOCO-like methods is the non-degenerate dependence of measurable parameters predicted by the model on the model parameters. Therefore an addition of different types of experimental data can noticeably improve the performance of such methods. For example, the trajectory response matrix is often expanded with the measured dispersion \cite{Sajaev:2005qw, Resende:2010cfa, Liu:2010ula, Benedetti:2011za, Martin:2014zpa} and betatron tunes~\cite{Smaluk:2016vgy}.

Inspired by the improvements of basic LOCO performance through the addition of dispersion and betatron tunes, further steps were made in terms of inclusion of various experimental data sets. For electron machines, for example, it is beneficial to use beam shape measurements with profile monitors based on synchrotron light. Other discussed data sets are the responses of closed orbit bumps to focusing field variations; betatron tune responses to focusing field variations; BPM-to-BPM betatron phase advances and beta functions in BPMs from turn-by-turn coordinates of kicked beam.

The development of the Sixdsimulation code was initiated as a tool for calculation of linear optics parameters in fully coupled circular machines in 6D phase space. Based on the developed analytical core, the code further evolved into a versatile tool with graphical user interface capable of highly automated linear lattice and closed orbit error measurements and corrections. The ease of integration into the control system of an accelerator, as well as the possibility to use saved data and other features allowed to use the code for the tuning of the VEPP-2000 collider, the VEPP-5 injector booster ring (both at BINP, Novosibirsk, Russia), and of the FAST linear accelerator (Fermilab, Batavia, Illinois).

In the present paper we describe the underlying principles of the expanded analysis algorithm (Sec.~\ref{sec:InvProbSolv}-\ref{sec:LatCorr}) and demonstrate the results of its application to various machines, including examples of possible ways to resolve degeneracies (Sec.~\ref{sec:ExpResults}).

\section{Inverse problem solver\label{sec:InvProbSolv}}
Both tasks of the closed orbit and linear lattice correction can be formulated as inverse problems, when some set of experimental data $V_{exp,j}$ is available, and the goal is to find the parameters $P_i$ of the model $\mathfrak{M}_j(P_i)$ that best describes the measurements. To find the approximate solution, the iterative method is used. The model predictions at the iteration $(n)$ are:
\begin{equation}
V_{mod,j}^{(n)}=\mathfrak{M}_j(P_i^{(n)}) \cdot s_j,
\end{equation}
where $s_j$ are normalization coefficients that can be used to modify the weights of some experimental data points. This approach is found to be very effective. For example, it can be used to freeze betatron tunes or improve the correction of the dispersion. Additionally, both $V_{exp,j}$ and $V_{mod,j}$ are assumed to be normalized to the statistical errors of $V_{exp,j}$.

The parameters of the model after iteration $(n)$ are:
\begin{equation}
P_i^{(n)} = P_i^{(0)} + \sum_{m=0}^{n-1} \Delta P_i^{(m)}.
\end{equation}

The difference between the experimental data and the model data is:
\begin{equation}
\label{difference_n}
D_j^{(n)} = V_{exp,j}-V_{mod,j}^{(n)}
\end{equation}

The goal is to find such variation of the parameters $\Delta P_i^{(n)}$ that cancels the residual difference between the model and the experimental data:
\begin{equation}
\Delta V_{mod,j}^{(n)} = -\Delta D_j^{(n)} = D_{j}^{(n)}.
\end{equation}

The model can be linearized in case of small parameter variation:

\begin{equation}
\begin{array}{lcl}

\Delta V_{mod,j}^{(n)} &=& s_j \left( \mathfrak{M}_j(P_i^{(n)}+\Delta P_i^{(n)}) - \mathfrak{M}_j(P_i^{(n)}) \right) \\
& \simeq &
s_j \left. \frac{\partial \mathfrak{M}_j}{\partial P_i}\right|_{P_i^{(n)}} k_i  \frac{\Delta P_i^{(n)}}{k_i} = \mathfrak{M}_{ji}^{(n)} \frac{\Delta P_i^{(n)}}{k_i}

\end{array}
\end{equation}

where $\mathfrak{M}_{ji}^{(n)}$ is the linearized and weighted model at iteration $(n)$:

\begin{equation}
\mathfrak{M}_{ji}^{(n)} = s_j k_i \left. \frac{\partial \mathfrak{M}_j}{\partial P_i}\right|_{P_i^{(n)}} .
\label{eq_linearModelMatrix}
\end{equation}

The variation of model parameters  can be obtained from Equation (\ref{eq_linearModelMatrix}) by applying pseudo inversion to the $\mathfrak{M}_{ji}^{(n)}$. Singular Value Decomposition (SVD) is a powerful method for such calculation. One of the remarkable features of this technique is easy control over the influence of the statistical error in the experimental data on the output result. Applying SVD gives the parameter variations at iteration (n):
\begin{equation}
\Delta P_i^{(n)} =k_i \sum_j \left( \mathfrak{M}_{ji}^{(n)} \right)^{-1}_{SVD} D_j^{(n)}.
\label{eq_parVarN}
\end{equation}
Summation over all iterations gives the total correction to the model parameters:
\begin{equation}
\Delta P_i = \sum_n k_i \sum_j \left( \mathfrak{M}_{ji}^{(n)} \right)^{-1}_{SVD} D_j^{(n)}.
\label{modelParsResult}
\end{equation}

To get the uncertainty of the $\Delta P_i$ its direct dependence from $V_{exp,j}$ is needed. It is more practical to search for direct dependence from $D^{(0)}_j$, which is, for this task is equivalent to $V_{exp,j}$, since $V^{(0)}_{j}$ does not depend on experimental data:
\begin{equation}
\Delta P_i = \sum_n \Delta P^{(n)}_i = T_{ij} D^{(0)}_j,
\end{equation}
here $T$ is final transformation matrix. Values $D^{(0)}_j$ are normalized to statistical sigmas, therefore uncertainty of reconstructed parameters is:

\begin{equation}
\sigma_{\Delta P,i} = \sqrt{ \sum_j T_{ij}^2 }.
\end{equation}

Induction method can be used to derive matrix $T$. For clarity both sets of normalization coefficients are set to unity. Equation (\ref{eq_parVarN}) gives dependence at step ``0'':

\begin{equation}
\Delta P^{(0)}_i=\left( \mathfrak{M}_{ji}^{(0)} \right)^{-1} D^{(0)}_j.
\end{equation}

For the next step equation (\ref{eq_parVarN}) need to be expanded to get direct dependence on $D^{(0)}_j$:
\begin{widetext}
\begin{equation}
\vspace{5mm}
\begin{array}{lcl}

\Delta P^{(1)}_i&=&\left( \mathfrak{M}_{ji}^{(1)} \right)^{-1} (V_{exp,j}-V^{(1)}_j) \\
&=& \left( \mathfrak{M}_{ji}^{(1)} \right)^{-1} (V_{exp,j}-\mathfrak{M}_{j}(P_i^{(0)}+\left( \mathfrak{M}_{ji}^{(0)} \right)^{-1}(V_{exp,j}-V^{(0)}_j))) \\
&\simeq& \left( \mathfrak{M}_{ji}^{(1)} \right)^{-1}(V_{exp,j}-\mathfrak{M}_{j}(P_i^{(0)}) -\mathfrak{M}_{ji}^{(0)}\left( \mathfrak{M}_{ji}^{(0)} \right)^{-1}(V_{exp,j}-V^{(0)}_j))\\
&=&\left( \mathfrak{M}_{ji}^{(1)} \right)^{-1}(I-\mathfrak{M}_{ji}^{(0)}\left( \mathfrak{M}_{ji}^{(0)} \right)^{-1}) D^{(0)}_j \\
&=& \left( \mathfrak{M}_{ji}^{(1)} \right)^{-1} \delta M_{ji}^{(0)} D^{(0)}_j.
\end{array}
\end{equation}
\end{widetext}

Resulting equation for matrix $T$ is:

\begin{equation}
T_{ij}=\sum_n \left( \mathfrak{M}_{ji}^{(n)} \right)^{-1} \delta M_{ji}^{(n-1)}\cdot\dots\cdot \delta M_{ji}^{(0)}.
\label{eq_difficultT}
\end{equation}

Equation (\ref{eq_difficultT}) may be very computationally intensive, since $\delta M_{ji}^{(m)}$ are square matrices with dimension $N\times N$, where $N$ is the number of experimental data. Recursive formula is more efficient:
\begin{equation}
\begin{array}{l}
T_{ij}=\sum_n T^{(n)}_{ij}=\\
=\sum_n \left(
\left( \mathfrak{M}_{ji}^{(n)} \right)^{-1} - \sum_{m=0}^{n-1}
\left( \mathfrak{M}_{ji}^{(n)} \right)^{-1} \mathfrak{M}_{ji}^{(m)}  T^{(m)}_{ij}
\right).
\end{array}
\label{eq_easyT}
\end{equation}

The $\chi^2$ gives an indication of the model quality:

\begin{equation}
\chi^2 = \left(\sum_{i} D_j^{(n)}\right)^2,
\end{equation}
If the systematic errors are absent, the equilibrium value of this function should be:
\begin{equation}
\chi^2_{eq} = N-M, \quad \sigma_{\chi^2_{eq}} = \sqrt{N-M},
\end{equation}
where $M$ is the number of the model parameters, and $\sigma_{\chi^2_{eq}}$ is the statistical sigma of the $\chi^2_{eq}$.

The fitted model is already a good result, but in most cases the goal is the correction of the found distortions. In the case of unsolvable errors, such as unexpected gradients in the main dipoles, another inverse task must be solved to find the new ideal configuration that accounts for the found features of the ring.

\section{Trajectory correction\label{sec:TrajCorr}}
One of the first tasks during machine commissioning is the trajectory correction. It consists of two parts: the first is to find distortions, and the second is to find corrector settings to compensate them.

It is possible to use calibrated BPMs to measure beam misalignments directly. However, the primary goal usually is to have the beam passing on the magnetic axis of the focusing elements. A beam based method can be used to solve this task. If the beam goes through the focusing element off axis, the change of its strength will modify the trajectory. This effect can be used to measure the beam position relative to the optical axis of the quadrupoles, solenoids, accelerating cavities and others.

Beam based trajectory measurement is an example of an inverse task. In this case, the model parameters $p_j$ are two shifts and two tilts of the relative trajectory at some point inside the element. The experimental data points $V_{exp,i}$ are responses of the trajectory in BPMs to the element strength variation. In general, the model has 4 adjustable parameters but for most short elements, such as quadrupoles, the tilts give too small of an effect to be reliably resolved. Matrix $\mathfrak{M}_{ji}$ is composed of trajectory responses in BPMs to base shifts of the trajectory. Due to the linear nature of the problem, only one iteration is needed to determine the relative trajectory.

For circular machines, the standard method is used to get the closed orbit distortion at the exit from an element:
\begin{equation}
X_{CO} = (I-M_{turn})^{-1}X_{0},
\end{equation}
where $X_{0}$ is the coordinate vector of the particle with zero initial displacement at the exit from the studied element. The 6D approach, in comparison to 4D, allows to automatically account for the effects of the energy deviation, since periodicity of the longitudinal degree of freedom follows from the used equation.

This method is not applicable for circular machines with tunes close to integer resonance. The preexisting distortion of the closed orbit will be changed due to the change of betatron tunes, introducing systematical errors in the measured responses.

During the correction, the directly or indirectly found distortions must be compensated for with selected set of correctors. This is again an inverse task. The variable model parameters $p_j$ are now the strengths of the correctors. Matrix $\mathfrak{M}_{ji}$ is composed of probe responses of the trajectory to the selected correctors. The flexibility of the SVD based method allows to control the balance between the correction quality and the strengths of the applied modifications.

The model, in this case, linearly depends on the parameters, but several iterations may be necessary if the theoretical closed orbit distortions from the correctors are used and they differ from the real ones.

\section{Linear lattice correction\label{sec:LatCorr}}

There are two different approaches to lattice correction. The first is aimed at achieving the best possible understanding of the studied accelerator and consequent correction, that often requires mechanical alignment of the elements. The second is a simplified approach suitable for fast routine use, and aimed to correct beta functions, dispersion, betatron tunes, coupling and other derived properties of the machine.

For the first approach, variable parameters of the model should be as complete as possible, including, for example, gradients in the main dipoles, quadrupole rotations and other parameters that cannot be corrected from the control room alone. After the lattice errors are measured, at least one of the following actions is required:
\begin{itemize}
\item Realign quadrupoles, correctors, BPMs, etc.
\item Adjust individual quadrupole currents.
\item Adjust the model if some errors are unremovable.
\end{itemize}

The faster method of lattice correction assumes that the variable parameters of the model contain only those that can be corrected immediately, for example, gradients corresponding to individual power supplies and various calibration coefficients. In this case, some residual systematic errors will remain but the resulting model will represent best fit of all inner imperfections with available knobs.

To decrease degeneracy and obtain the best understanding about the real lattice configuration, it is beneficial to collect as much experimental data as possible. Further in this section, all data sets implemented in "Sixdsimulation"  will be discussed.

\subsection{Closed orbit responses to dipole correctors}
This is the most basic set of experimental data that can be measured naturally in most accelerators. It provides a good overall signal-to-noise ratio and allows to reconstruct beta functions and phase advances between BPMs and correctors.
\subsection{Dispersion}
If directly measured and included into experimental data, dispersion can be corrected to the level of measurement precision or to the level allowed by the flexibility of machine adjustments knobs. If not included, it is prone to relatively high residual errors, since it is indirectly connected to the closed orbit responses.
\subsection{Betatron tunes}
Since betatron tunes are just full machine phase advances, they can be reconstructed from the orbital responses with good precision. In most machines, it is possible to measure and correct betatron tunes precisely before measuring experimental data. Known betatron tunes could be used in several ways.

First, it might be beneficial to exclude tunes from experimental data and use them for independent cross check of fitted lattice. Second, if corrected to the model values, tunes could be included into the experimental data set with artificially small sigmas in order to stabilize the fitting algorithm, by prohibiting the use unrealistic distortions. In the case of big initial distortions, it will also give a smaller shift of the working point after the correction implementation. And, lastly, the tunes can be simply included into the experimental data set as a usual parameter.

\subsection{Betatron tune responses to focusing elements strength variations}
If the machine automation system allows precise betatron tune measurements, it could be useful to extend the experimental data set with betatron tune responses to focusing element strength variations.

Let us consider uncoupled motion in one plane, where $\mu_0$ and $\Delta \mu$ are the total phase advance and its variation due to a small change in the thin focusing element with integrated strength $\Delta F$. If the variations are small:

\begin{equation}
\Delta \mu \ll 1 \quad \mathbf{and} \quad  \Delta \mu \ll \tan \mu_0,
\end{equation}
then phase advance variation can be simplified to:
\begin{equation}
\label{tuneVar}
\Delta \mu = 2\pi\Delta\nu=\frac{\Delta F \beta_{el}}{2},
\end{equation}
where $\beta_{el}$ is the beta function in the varied element.

Skew-quadrupoles can also be used as a source of tune variations, especially if fractional parts of tunes are close to each other. For a non-resonant case, a variation $\Delta S$ of integrated strength in a thin skew-quadrupole gives:
\begin{equation}
\label{eq_dnuFromSQdG}
\delta \mu_x=-\frac{1}{4}\frac{\beta_x\beta_y\sin\mu_y}{\cos\mu_x-\cos\mu_y}\Delta S^2;
\quad\delta \mu_y = -\delta \mu_x\frac{\sin\mu_x}{\sin\mu_y} .
\end{equation}

For a resonance case of $\cos\mu_x=\cos\mu_y$, even a small skew-quadrupole produces full coupling with indistinguishable modes. For a thin lens approximation the coupling parameter (tune half splitting) is:
\begin{equation}
\label{eq_dnuFromSQdG_res}
\delta\mu=\sqrt{\beta_x\beta_y} \Delta S,
\end{equation}
with equal excitation of both transverse modes.

\subsection{Beam shapes}
If the particles emit enough synchrotron radiation in bending magnets, optical imaging systems can be used as precise beam position monitors. In addition, if the transverse sizes of the beam are much bigger than diffraction limits, cameras also give reliable information about transverse shape of the beam. The set of measured second moments of transverse beam distribution gives a simple way to expand experimental data with precise values reflecting transverse beta functions, emittances and coupling parameters. Similarly to betatron tunes, beam shapes can be excluded from fitting and used to crosscheck the adjusted model.

\subsection{Closed orbit bumps responses to focusing strength variations}
Valuable set of experimental data could be obtained by measuring closed orbit distortion variations in response to the focusing variation of individual elements or families. To measure such a response one needs to combine four closed orbit measurements:
\begin{enumerate}
\item The initial closed orbit $V_0$.
\item The closed orbit distorted with some dipole corrector $V_b$.
\item The closed orbit with correctors and focal strength variation $V_{bf}$.
\item The closed orbit with only focal strength variation $V_f$.
\end{enumerate}

The combination:
\begin{equation}
V_{resp}=(V_{bf}-V_f)-(V_b-V_0)
\end{equation}
gives the closed orbit bump response to the focal strength variation separated from the initial response of the closed orbit.

One of the main drawbacks of this method is low signal-to-noise ratio of the data. Let us estimate the typical value of response in a simplified approach, using the following assumptions:
\begin{itemize}
\item No transverse coupling.
\item All elements and correctors are thin, and the varied element is a normal quadrupole.
\item Bumps are created by a single corrector with fixed invariant excitation $A_{corr}$.
\item There are many correctors, so that it is always possible to chose one that creates the biggest orbit distortion in the varied element for a chosen invariant excitation value $A_{corr}\sqrt{\beta_{el}}$.
\item Variation of phase advance across the ring from the element's focusing strength variation is small ($\Delta \mu \ll 1$ and $\Delta \mu \ll \tan \mu_0$).

\end{itemize}

For a given orbit distortion of $A_{corr}\sqrt{\beta_{el}}$ at the element location, the element's focusing strength variation will induce a trajectory kick:
\begin{equation}
\Delta x'_{el}=\Delta F A_{corr}\sqrt{\beta_{el}}.
\end{equation}
The kick will create a closed orbit distortion:
\begin{equation}
x_{el}(s)=\frac{\sqrt{\beta(s)}}{2 \sin \pi \nu} \Delta x'_{el} \sqrt{\beta_{el}}\cos \left(\left|\psi(s)-\psi_{el}\right|-\pi \nu\right).
\end{equation}
Considering eq. (\ref{tuneVar}) this corresponds to the excitation of orbit distortion with invariant $A_{el}$:
\begin{equation}
A_{el}=A_{corr}\frac{2\pi \Delta \nu}{\sin \pi \nu}.
\end{equation}

Usually, tune variation can not exceed several units of $10^{-2}$, and  $\sin \pi \nu$ lies between 0.1 and 1, therefore the bump variation is about 10 times smaller than responses to correctors. Implementation of local bumps could be beneficial, but it is not always possible.

For some cases the use of this data is highly recommended. First, if there is a region with large phase advance ($\Delta\psi \gtrsim \pi$) without correctors and BPMs it might be impossible to locate a source of distortion. For example, the interaction region in a collider is a particularly difficult case, because of the phase advance between the final focusing groups being close to $\pi$.

Another valuable outcome from the use of such data is the calibration of focusing elements. This is especially true for skew-quadrupoles, since the tune responses are insensitive to the polarity of skew-gradient variation. In the non-resonant case the tunes depend quadratically on skew-gradient variation (\ref{eq_dnuFromSQdG}), and in the resonant case the modes are excited equally (\ref{eq_dnuFromSQdG_res}).

Finally, each individually powered quadrupole will effectively give data equivalent to weak vertical and horizontal correctors with the same rotational error as parent quadrupole. Measurement of bump responses uses the same instrumentation as that needed for measuring  standard LOCO data set an thus requires only additional time.

\subsection{BPM-to-BPM betatron phase advances and beta functions in BPMs}
Analysis of turn-by-turn BPM data of the kicked beam can provide both BPM-to-BPM betatron phase advances and beta functions in BPMs, all arbitrary scaled to the same coefficient \cite{Huang:2005gd}.

With enough BPMs this data set alone can be used to directly describe errors of the accelerator lattice. Lattice fit procedure can be used in this case to calculate corrections necessary to bring accelerator to the design configuration. If used alone, this data cannot provide reliable information about quadrupole gradients errors for lenses that form groups of 5 and more in between BPMs.  Using this data set in conjunction with standard LOCO will improve beta function reconstruction in the areas close to BPMs.

\section{Experimental results\label{sec:ExpResults}}
\subsection{VEPP-2000}

VEPP-2000 collider was designed for the refinement of the cross section of the $e^+e^-$ annihilation to hadrons in the energy range $0.4\div2$~GeV \cite{Shatunov:2000zc}. VEPP-2000 is a small ring with about 24~m circumference operated in the single bunch mode. The concept of round colliding beams~\cite{Danilov:1996jw} is used to achieve high luminosity in two head-on collision points.
The linear lattice is formed by 8 identical dipoles, 24 quadrupoles grouped in 6 families and 4 superconducting solenoids of the final focus regions. The ring has two-fold symmetry, with the exception of the RF cavity and the solenoidal field in one of the particle physics detectors (CMD), which is significant only at the beam energy below 300 MeV.

Beam diagnostics used for lattice corrections consists of 4 electrostatic BPM pickups, and 16 digital cameras that register synchrotron radiation. The pickups can measure turn-by-turn and averaged beam coordinates. The cameras provide relative beam positions and shapes, absolute position measurements are possible only after beam based calibration.

In order for the round colliding beams concept to work, the linear lattice should be precisely tuned. Table \ref{t_v2k_parameters} contains major parameters of VEPP-2000 lattice and their tolerances. The energy-scanning mode of operation of VEPP-2000 requires frequent energy changes, each introducing poorly predictable distortions:
\begin{itemize}
\item Solenoidal field of the CMD detector is not scaled with energy, which leads to energy-dependent lattice configurations, with bigger changes at lower energies.
\item Small variations of the closed orbit lead to significant distortions of the linear lattice due to nonlinear fields.
\item Iron saturation effects are significant at higher energies.
\end{itemize}

\begin{table}[bth]
\centering
\caption{Major lattice parameters and their tolerances.}
\begin{tabular}{|l|c|c|}
\hline
Parameter & Nominal value & Max. error  \\ \hline
$\beta_x$ and $\beta_y$ at the IP  & 5$\div$8 cm & 5\%  \\ \hline
$\beta_x$ and $\beta_y$ in the arcs  & 20$\div$500 cm & 10 \%  \\ \hline
Horizontal betatron tune, $\nu_x$ & 4.1$\div$4.2 & 0.001 \\ \hline
Vertical betatron tune, $\nu_y$ & 2.1$\div$2.2 & 0.001 \\ \hline
Disp. at the IP, both planes & 0 cm &  1 cm\\ \hline
Horizontal disp. in the arcs, $D_x$ & 0$\div$100 cm &  5 cm\\ \hline
Vertical disp. in the arcs, $D_y$ & 0 cm &  5 cm\\ \hline
\end{tabular}
	\label{t_v2k_parameters}
\end{table}

\begin{figure}[htb]
    \centering
	\includegraphics[width=0.9\columnwidth]{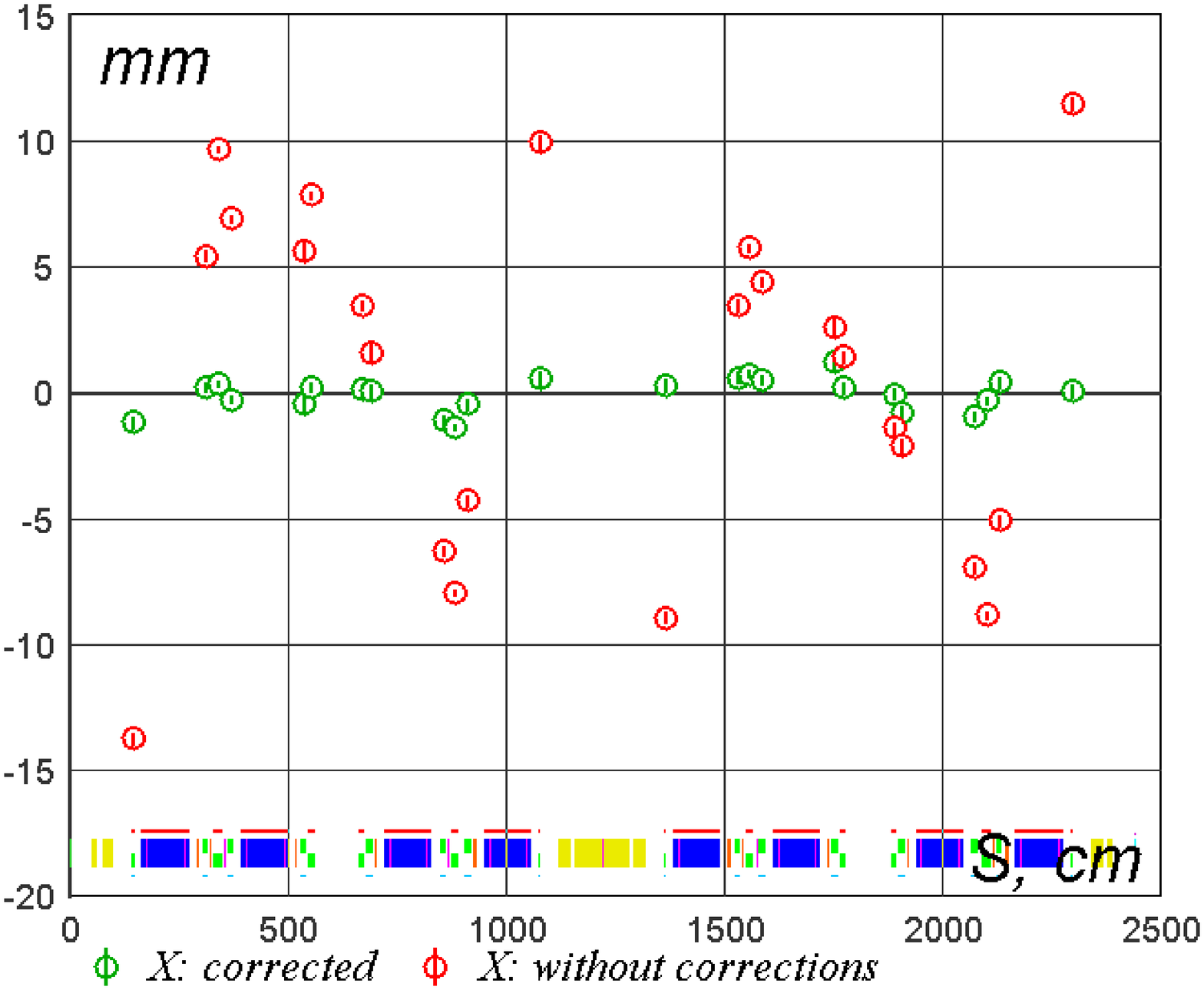} \\
	\includegraphics[width=0.9\columnwidth]{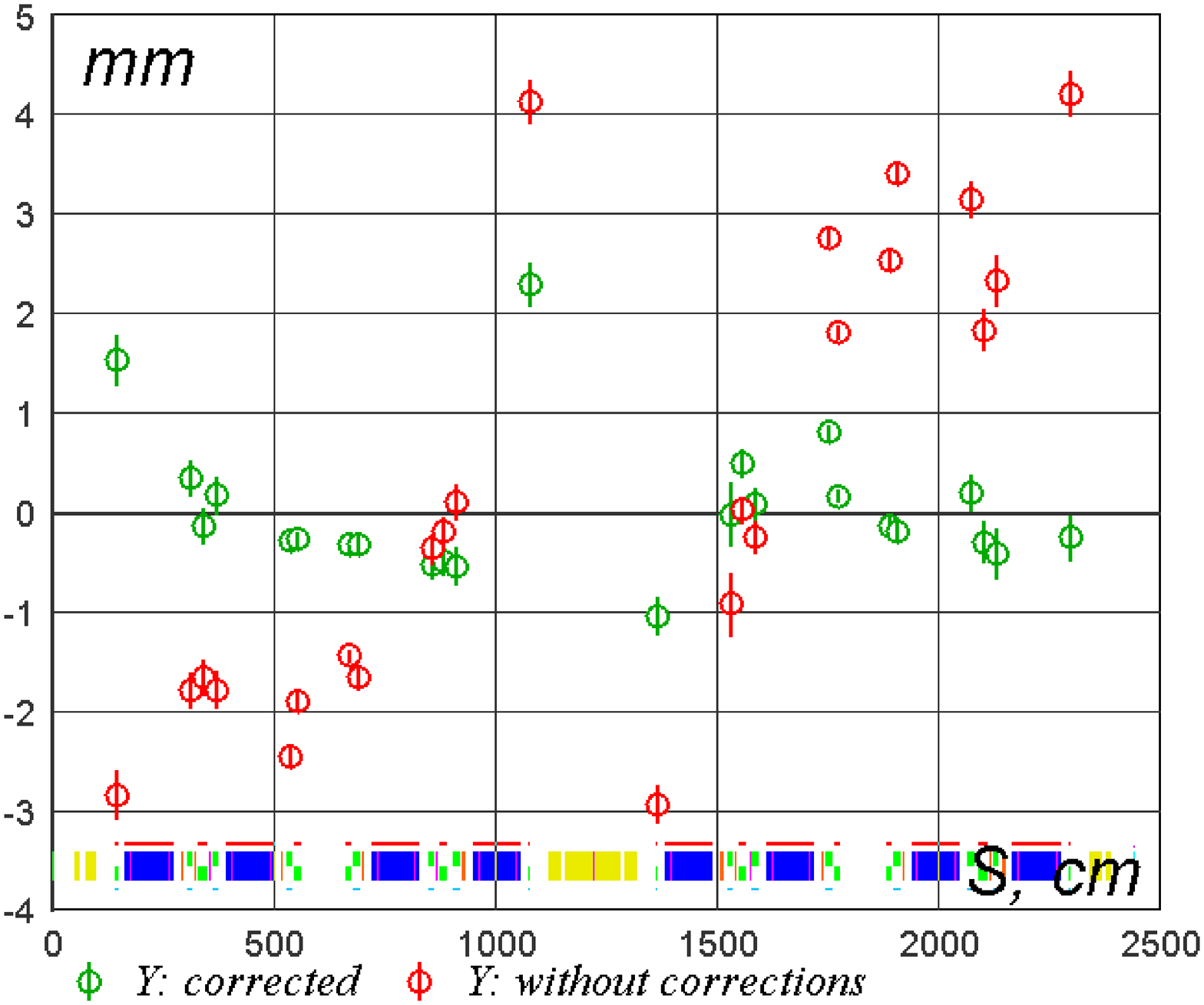}
    \caption{Closed orbit measured in quadrupoles with and without corrections (green and red) for horizontal plane (left) and vertical plane (right) in VEPP2000 collider at the energy of 240~MeV per beam.}
    \label{fig:v2kOrbCorr}
\end{figure}

\begin{figure}[bt]
    \centering
	\includegraphics[width=0.9\columnwidth]{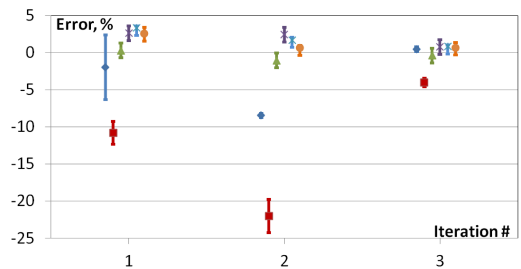}
    \caption{Changes in excitation currents averaged across the families of quadrupoles in three consequent corrections in percent; error bars show average deviations of corrections for the quads in the same family $\left<\left|\left<\Delta I\right>-\Delta I\right|\right>/I_0$ }
    \label{fig_v2k_deltas123}
\end{figure}

After each energy change the closed orbit and lattice should be tuned to get the best performance of the collider. Therefore robust correction tools are important for the successful operation of VEPP-2000.

The main source of the closed orbit distortion at VEPP-2000 are the misalignments of the final focus solenoids. The proper alignment is very difficult due to the complex structure of the cryostat and significant coil-to-coil forces that can shift solenoid components. Also no individual field measurements were done after full assembly.

\begin{figure}[tb]
\centering
\includegraphics[width=\columnwidth]{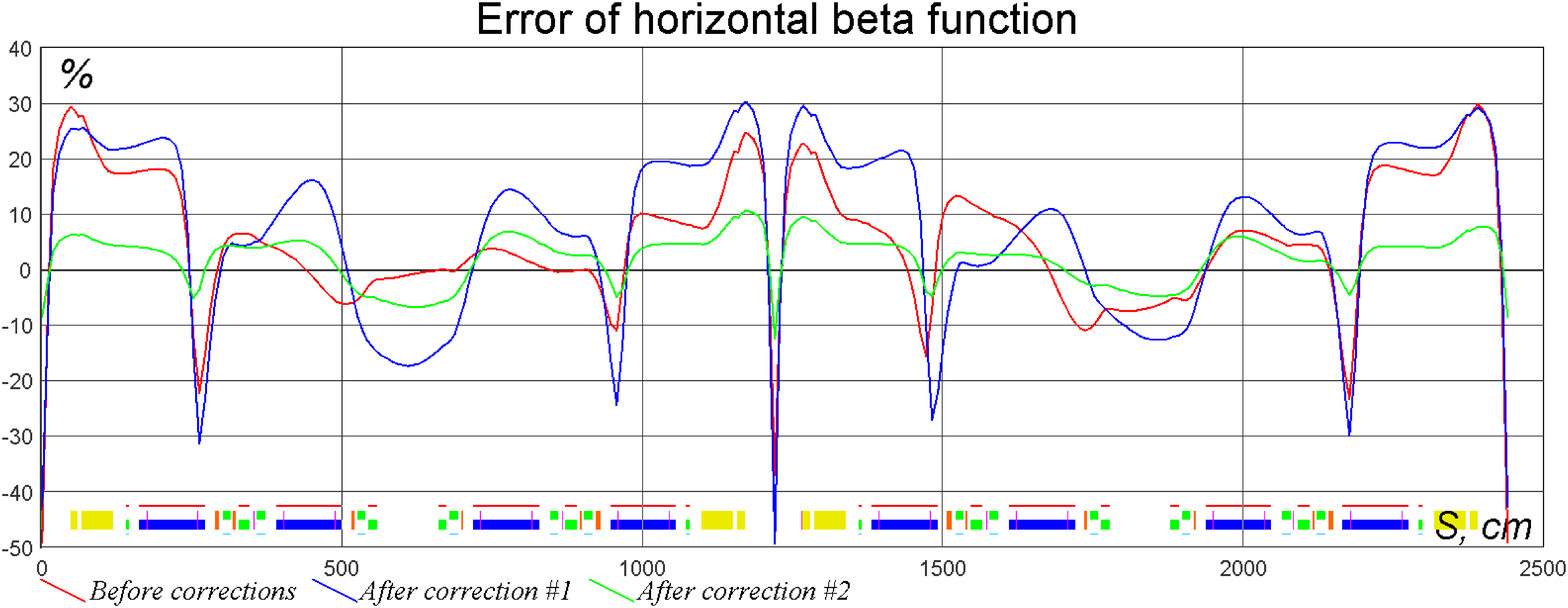} \\
\includegraphics[width=\columnwidth]{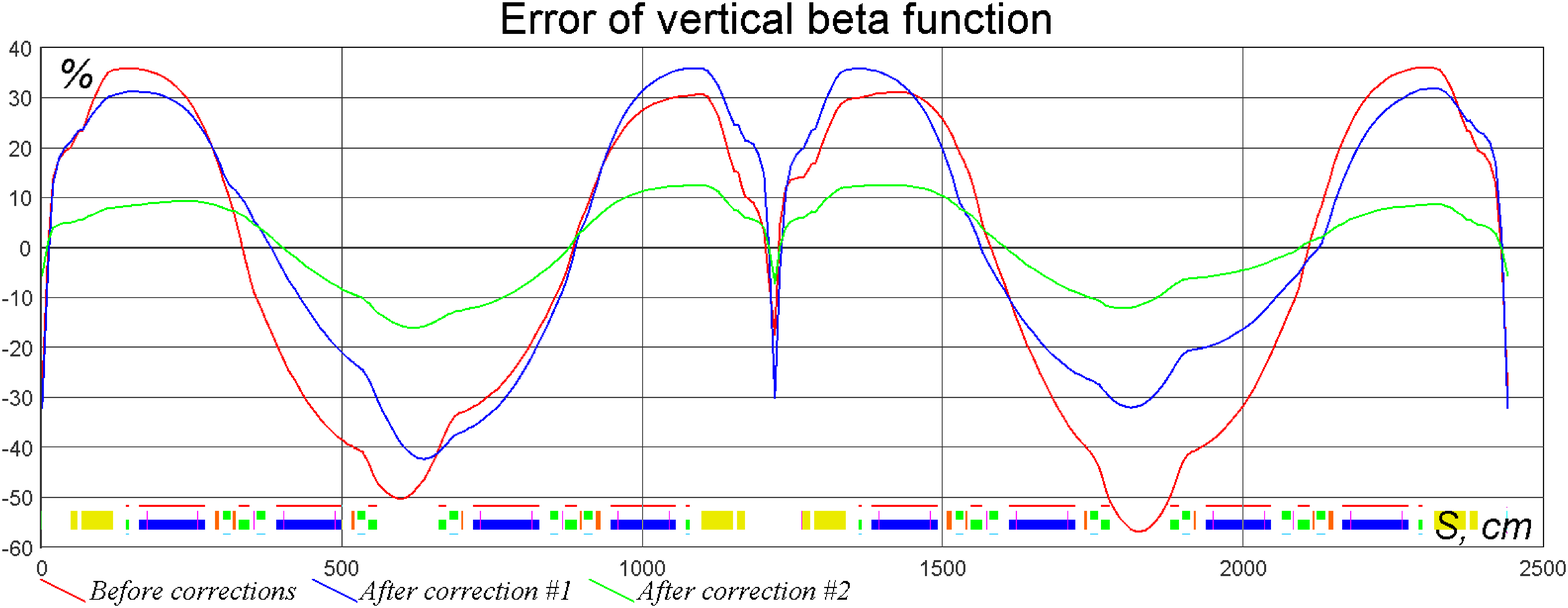} \\
\includegraphics[width=\columnwidth]{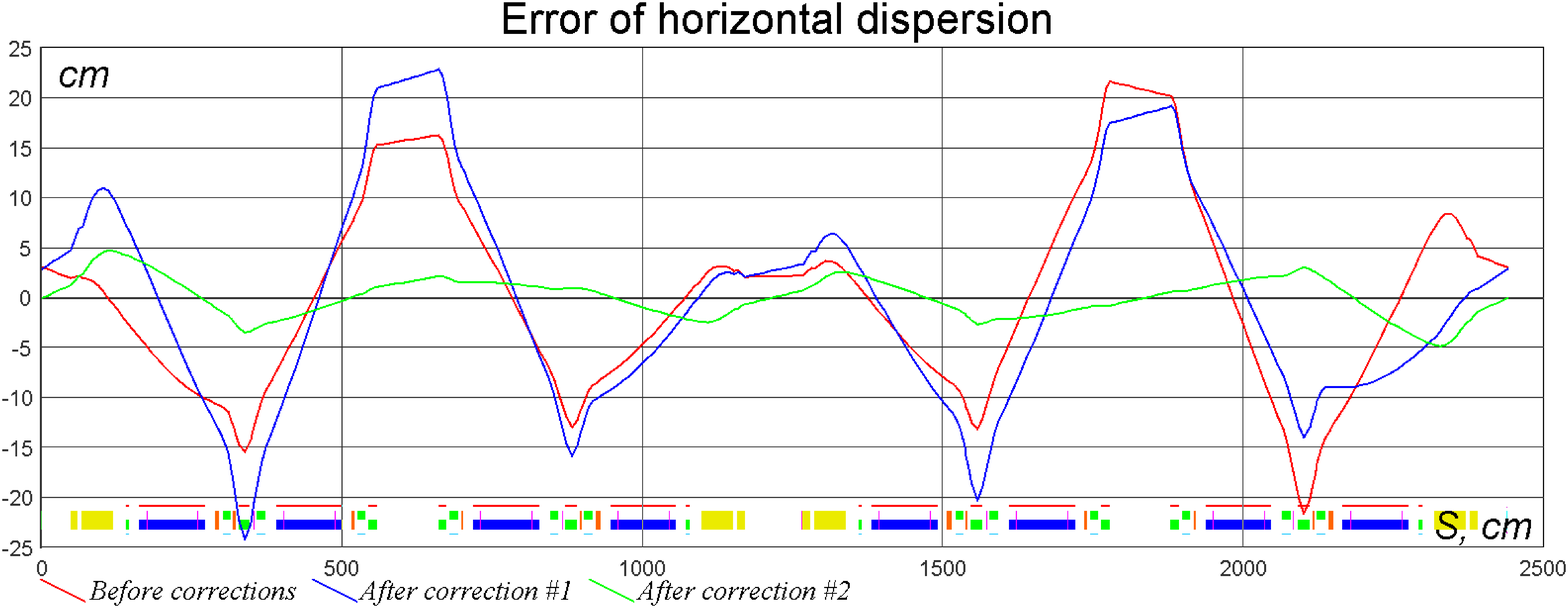}
\caption{Horizontal beta functions, vertical beta functions and horizontal dispersions in models fitted to match three consequent experimental data sets during VEPP-2000 lattice correction at 900 MeV}
\label{f_v2k_flockeCorrection}
\end{figure}
\afterpage{\clearpage}

\begin{figure}[tb]
\centering
\includegraphics[width=\columnwidth]{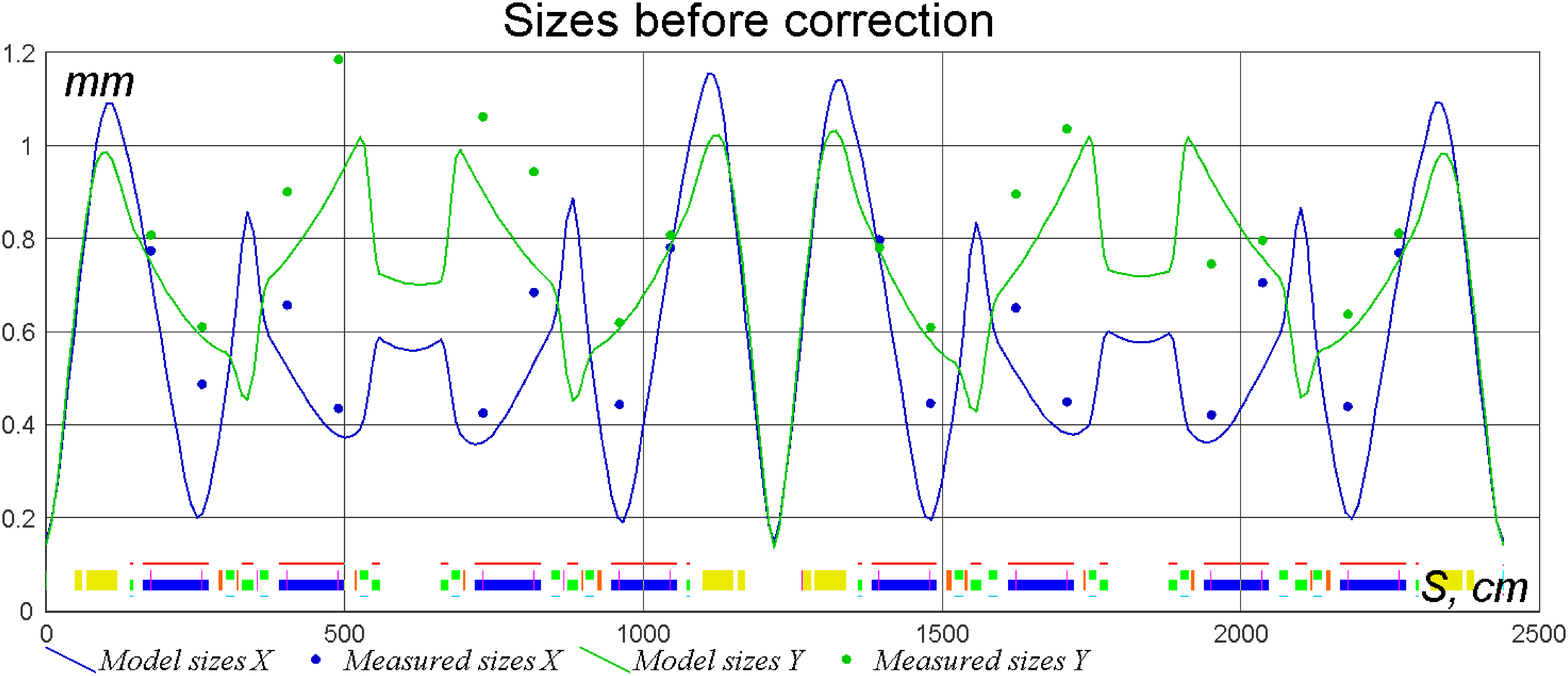} \\
\includegraphics[width=\columnwidth]{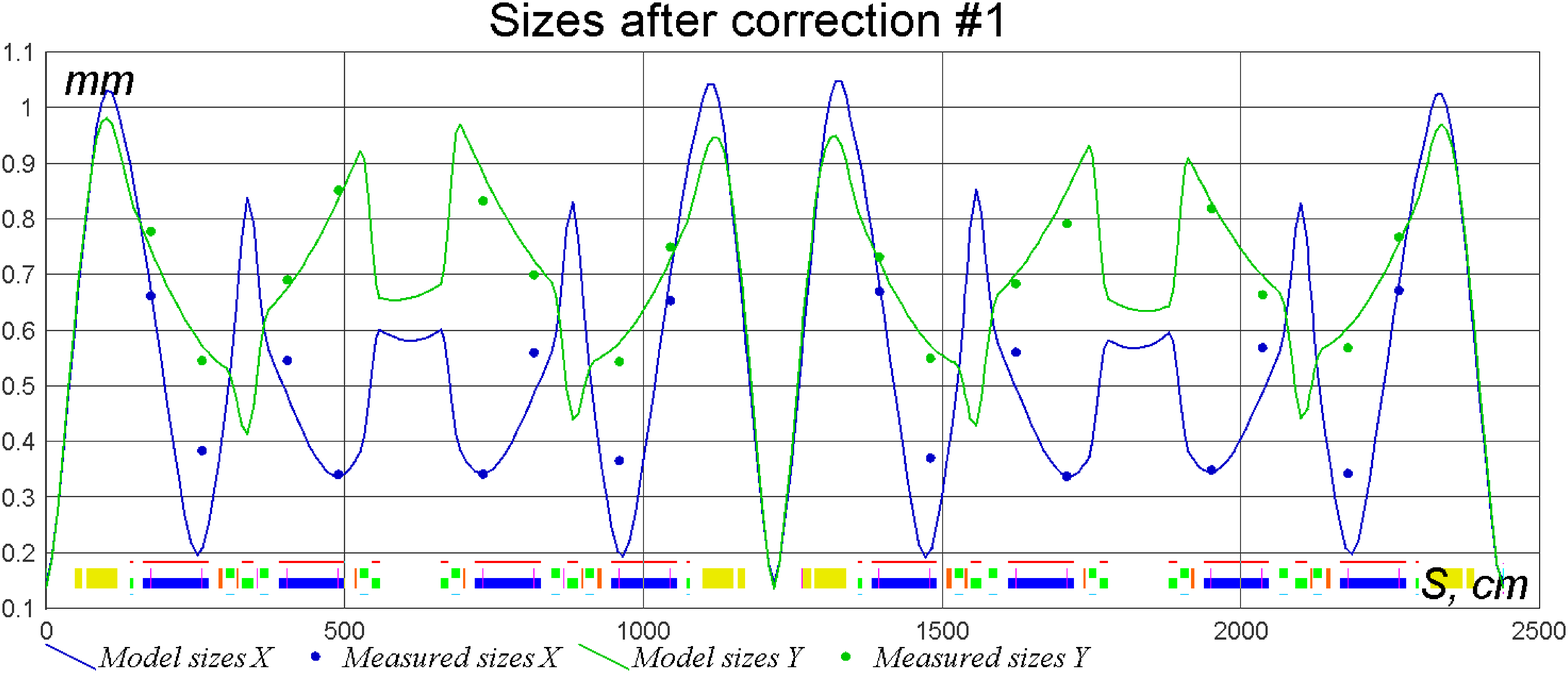} \\
\includegraphics[width=\columnwidth]{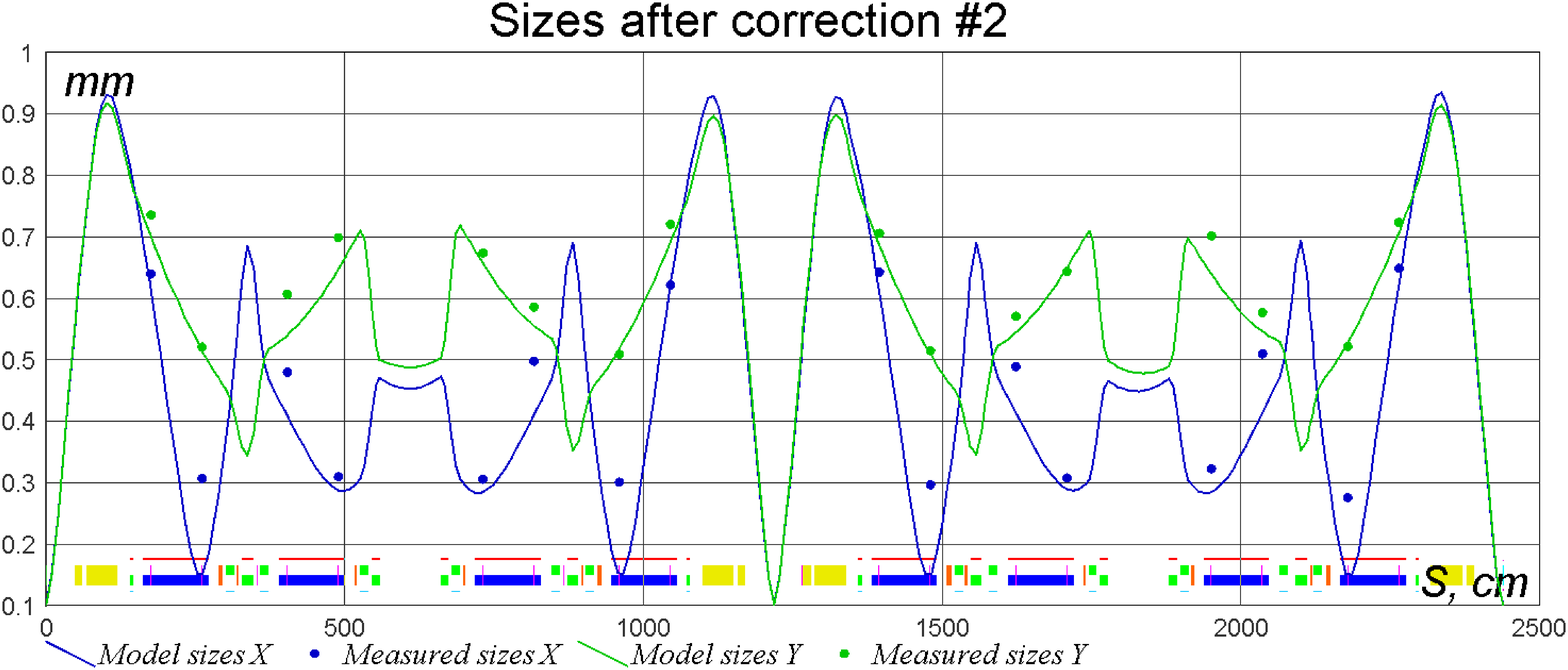}
\caption{Comparison of model and measured beam sizes before correction, after first correction and after second correction. Errors of the measured sizes are less then size of the dot}
\label{f_v2k_sizesCorrection}
\end{figure}
\afterpage{\clearpage}

The main task of the orbit correction in the VEPP-2000 is to align the closed orbit as close as possible to the magnetic axes of the elements. After achieving the best orbit, one can store ideal positions of the beam at the locations of the BPMs and use them for future orbit corrections. Plots in Figure \ref{fig:v2kOrbCorr} show the closed orbit in quadrupoles with and without correction.

After good closed orbit correction is done, the next step in order to reach the best performance of the collider is the precise linear lattice correction.

One of the typical examples of critical necessity to use the advanced correction methods for the VEPP-2000 collider ring was the operation at the top beam energy of 900~MeV. Saturation effects in warm magnets caused significant distortion of lattice with beta functions being off by about one hundred percent. The correction with three iterations was implemented. The superconducting solenoids were already well calibrated and therefore were excluded from the set of variable parameters of the model. Figure \ref{fig_v2k_deltas123} shows corrections to currents for different families of quadrupoles in three consequent iterations. Figure \ref{f_v2k_flockeCorrection} shows normalized errors of beta functions ($100\times(\beta_{model}-\beta_{fit})/\beta_{model}$) and error of dispersion before corrections and after the first and second corrections. Beam sizes were excluded from experimental data for crosscheck, Figure \ref{f_v2k_sizesCorrection} shows measured sizes compared to beam envelopes calculated from fitted models.

During the first correction, variations of the model were intentionally limited and therefore the fitted structure did not reflect actual condition of the accelerator. The second and third fits were not limited and demonstrated better agreement with the measured beam sizes. Convergence of excitation current corrections as well as increased luminosity also confirmed successful optimization of lattice parameters.

\begin{figure}[tb]
\centering
\includegraphics[width=0.9\columnwidth]{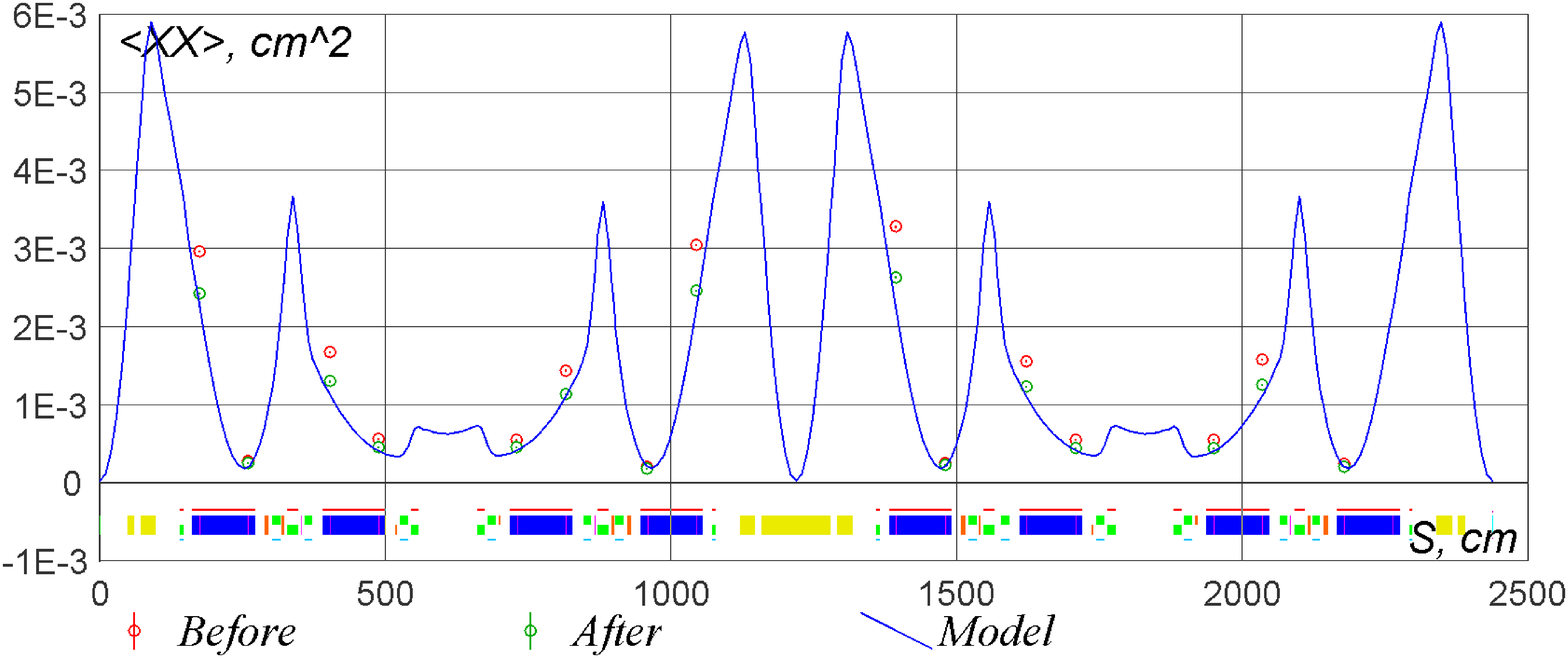} \\
\includegraphics[width=0.9\columnwidth]{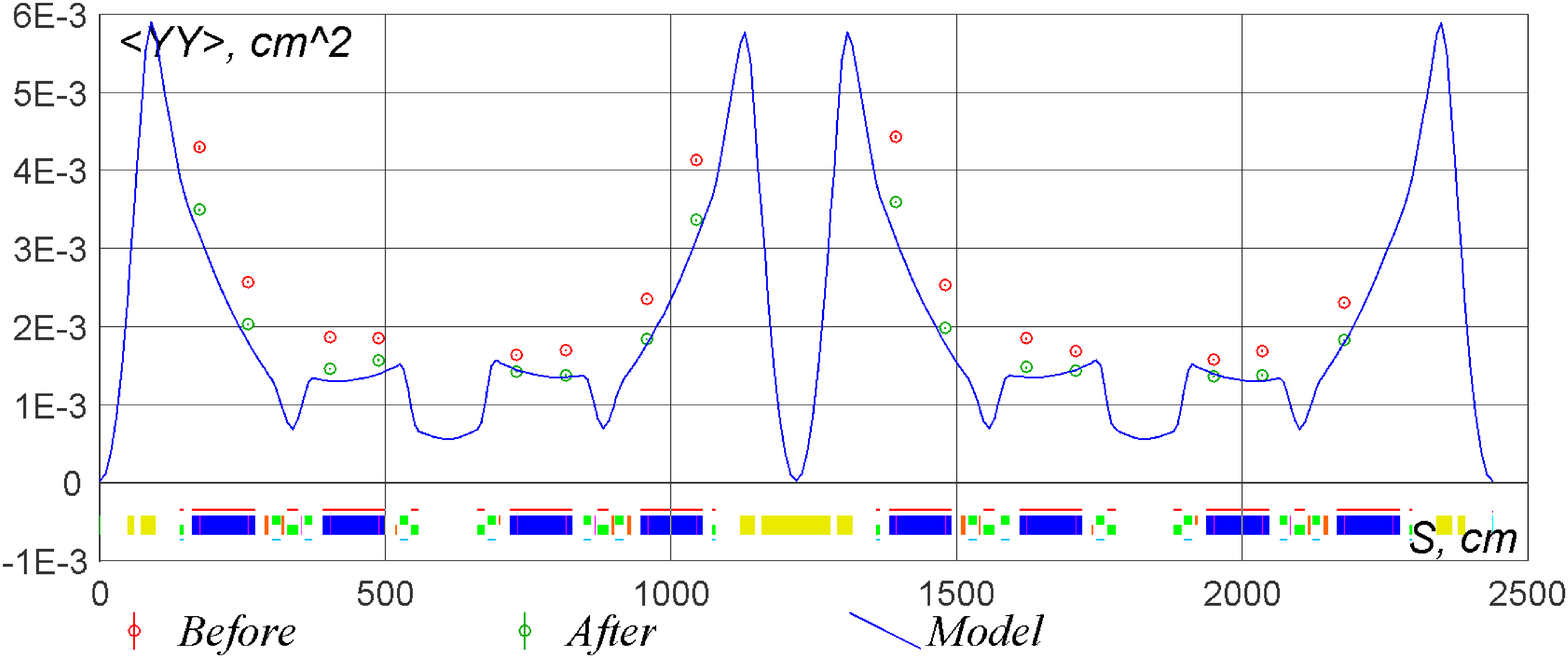} \\
\caption{Measured second moments before correction (red circles), and after two corrections based on beam shapes (red circles)}
\label{fig_v2k_SMCorrection}
\end{figure}

As it was mentioned above, VEPP-2000 has 16 digital cameras to measure both position and transverse shape of circulating beams. This means that there are, on average, about 4 and 8 measurements per $2 \pi$ of betatron phase advance in horizontal and vertical planes. Therefore LOCO data set can be substituted with measurements of second moments. Measurements of beam shapes take just a few minutes compared to about 20-30 minutes for LOCO data set. It was found, that for VEPP-2000 first iterations of lattice correction can be done with experimental data composed of betatron tunes, dispersion and second moments, and only one final iteration is preferred to be based on data that includes responses to dipole correctors. Figure \ref{fig_v2k_SMCorrection} shows second moments before and after 2 iterations based on beam transverse shapes.

\begin{figure}[bt]
\centering
	\includegraphics[width=0.9\columnwidth]{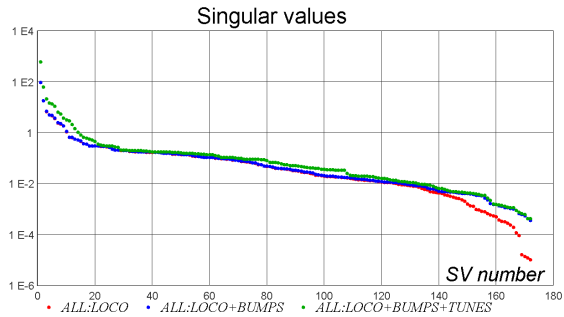}
    \caption{Spectrum of singular values for all fitting parameters of VEPP-2000 for various data sets: LOCO data set with dispersion and betatron tunes (red); with the addition of bump responses to focusing strength variation (blue); with the addition of tune responses to focusing strength variations (green). }
    \label{fig:v2kAllSVSpectr}
\end{figure}

Even though a good quality lattice was possible to achieve for the VEPP-2000 using standard LOCO data set expanded with dispersion and betatron tunes, some degeneracy still remains in the parameters space. For example, this data set is weakly sensitive to symmetric variations of solenoidal fields in the solenoids of final focuses. Degenerate degrees of freedom cannot be corrected by the standard method, and thus the entire correction procedure relies on having small initial errors associated with the degenerate degrees of freedom. Degeneracy shows itself as small singular values. Therefore, the modified data set with resolved degeneracy should have raised the exponential tail in the spectrum of singular values.

Additional data sets mentioned in section \ref{sec:LatCorr} were implemented in "Sixdsimulation" solver with the aim to reduce the effect of big initial errors in degenerate degrees of freedom. Figures \ref{fig:v2kAllSVSpectr} show the impact of additional data sets on the specter of singular values for all fit parameters:
\begin{itemize}
\item Standard LOCO, dispersion, betatron tunes.
\item Same as above plus bump responses to focusing strength variations.
\item Same as above plus betatron tune responses to focusing strength variations.
\end{itemize}

It is obvious that additional data reduced the degeneracy, for some singular values by as much as a factor of 100. The remarkable improvement for the four smallest singular values correspond to the resolution of degeneracy resulting from the similarity of the effects of *D1 quadrupole rotation and gradient in the adjacent skew-quad *SQ2, where * stands for the number of one of the four families. Option to use additional data sets was implemented during last shutdown of VEPP-2000 and will be tested experimentally later.

\subsection{Damping ring of the VEPP-5 injection complex}

Reliable operation of the injection complex VEPP-5 is crucial for both VEPP-4m and VEPP-2000 colliders at BINP, since it is the only source of positron and electron beams for these machines \cite{Starostenko:2014iaa}. One of the key components of the injection complex is the damping ring. The ring has two fold symmetry and is formed by 8 identical dipoles and 28 quadrupoles grouped in 7 families.  Beam based corrections were applied at damping ring to better understand its configuration.

\begin{figure}[htb]
\centering
	\includegraphics[width=0.9\columnwidth]{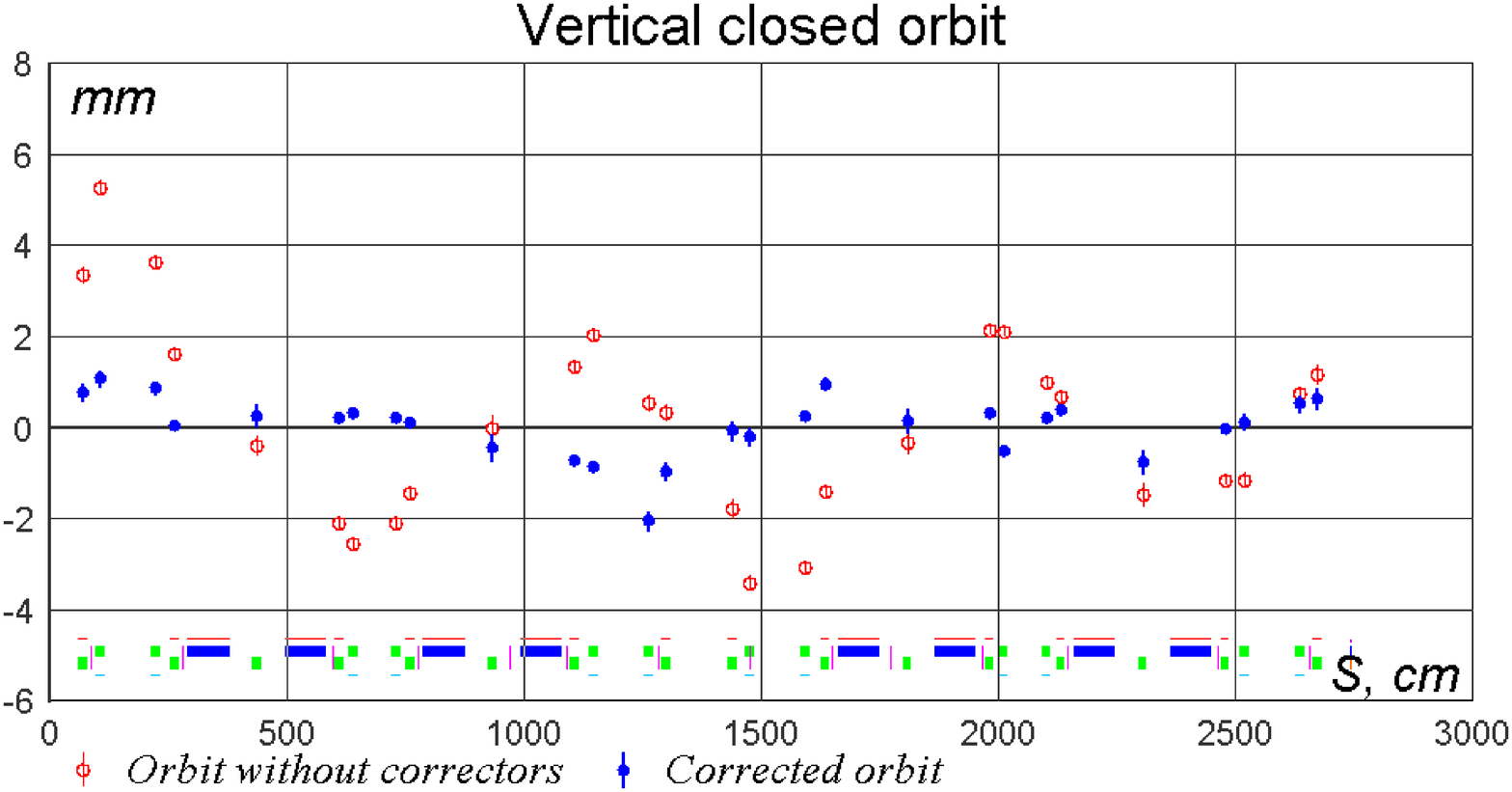}
    \caption{Closed orbit before (red circles) and after correction (blue dots) for vertical plane in damping ring of VEPP-5 injection complex}
    \label{fig:v5OrbCorr}
\end{figure}

First, manual corrections were done, to ensure correctness of commutations of BPMs and correctors, approximate calibrations, betatron tunes and some other parameters.
Rough setup of lattice allowed to use beam based method of closed orbit alignment along quadrupoles' axes. Figure \ref{fig:v5OrbCorr} illustrates correction of closed orbit in quadrupoles on an example of vertical orbit.

\begin{figure}[htb]
\centering
	\includegraphics[width=0.9\columnwidth]{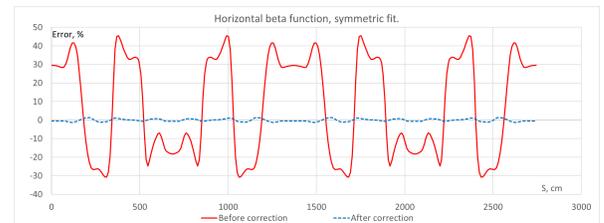}
    \caption{Horizontal beta function in fitted models before and after corrections. All elements from the same family assumed to be identical.}
    \label{fig:vepp5LatCorrBx}
\end{figure}

\begin{figure}[htb]
\centering
	\includegraphics[width=0.9\columnwidth]{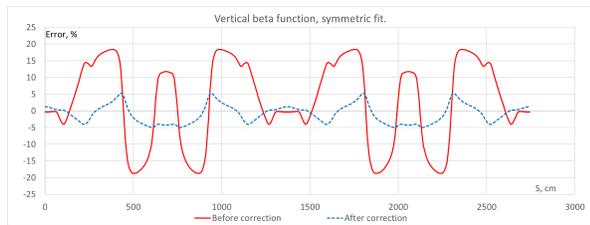}
    \caption{Vertical beta function in fitted models before and after corrections. All elements from the same family assumed to be identical.}
    \label{fig:vepp5LatCorrBy}
\end{figure}

Lattice corrections were done with 14 BPMs with about 100 $\mu m$ resolution, 20 horizontal and 12 vertical correctors. In order to stabilize fitting, all elements in the families were assumed to be identical. Later assumption were applied to quadrupoles, calibrations of correctors and calibrations of BPMs. Due to the absence of skew-qudrupoles in the ring, coupling effects, such as vertical dispersion, vertical orbit responses to horizontal correctors and vice versa were disregarded. Resulted optical functions for fitted lattices before and after corrections are presented in Figures \ref{fig:vepp5LatCorrBx} and \ref{fig:vepp5LatCorrBy}.

\subsection{FAST}
\begin{figure}[htb]
	\includegraphics[width=\columnwidth]{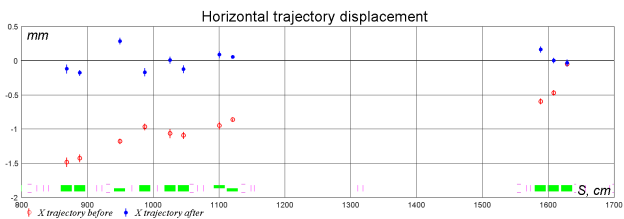}
    \caption{Horizontal trajectory offsets in quadrupoles in FAST linear accelerator before and after correction (red and blue circles)}
    \label{f_FAST_trajCorr}
\end{figure}

The low energy part of the FAST linear accelerator based on 1.3 GHz superconducting RF cavities was successfully commissioned at Fermilab \cite{FAST}. After the initial setup of FAST was completed for the 2016 run the model-dependent orbit correction relative to the quadrupole magnetic axes was done, resulting in less than 500~$\mu$m errors in both planes.  Figure \ref{f_FAST_trajCorr} illustrates trajectory correction at the FAST on an example of the horizontal plane.

Lattice correction was done for the real experimental setup of capture cavities and quadrupoles. First, the parameters of focusing elements were fit by analyzing trajectory responses to the dipole correctors. Then, initial conditions were reconstructed with the help of beam second moments measured at several beam profile monitors located along the FAST beam line.

\begin{figure}[htb]
    \centering
	\includegraphics[width=\columnwidth]{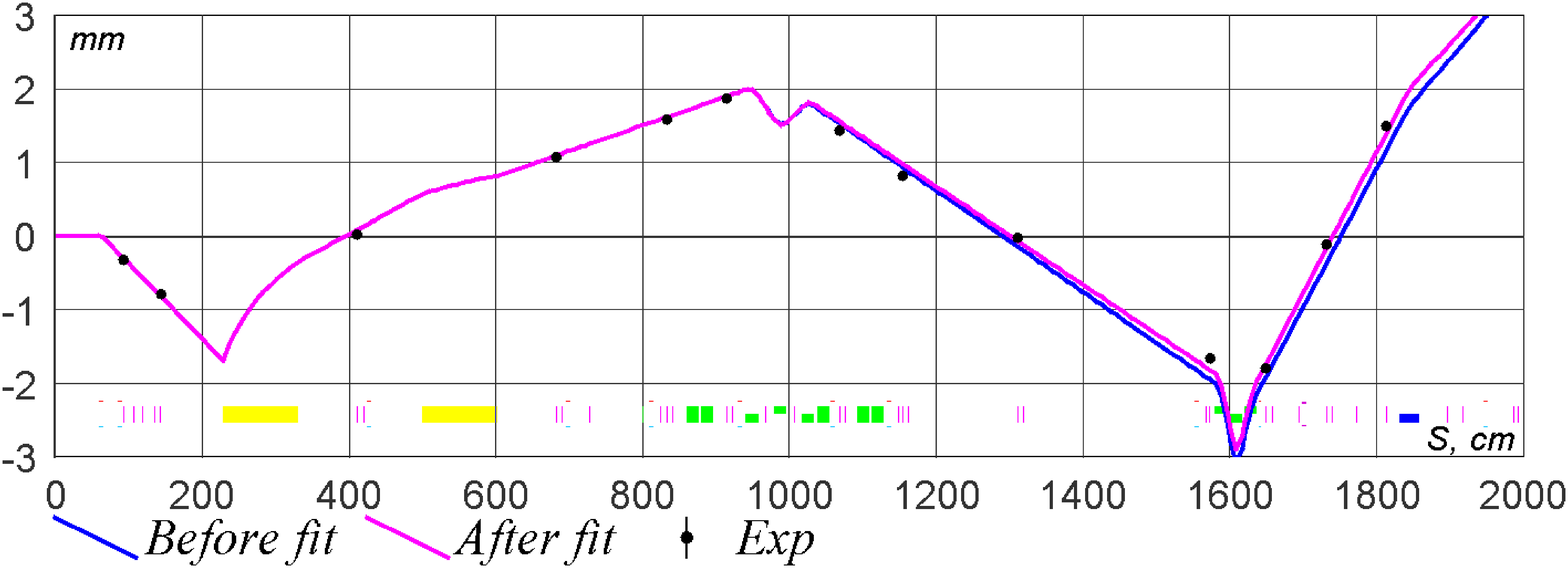}
	\includegraphics[width=\columnwidth]{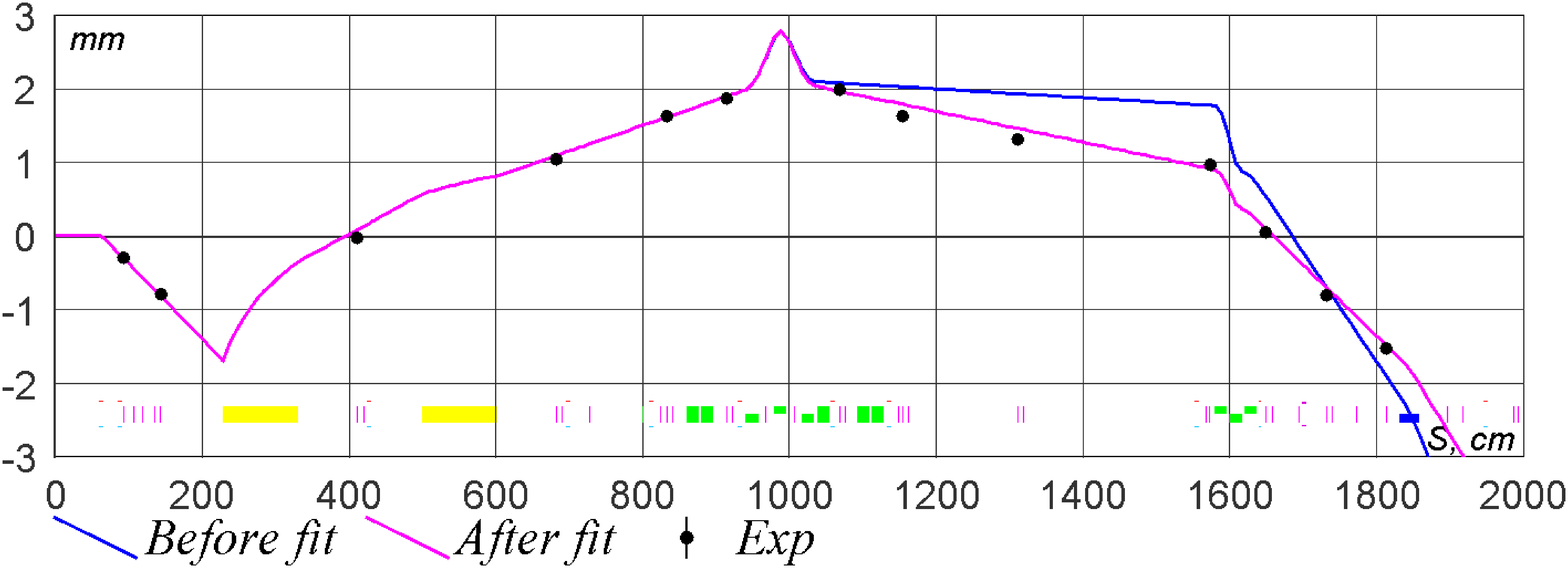}
    \caption{Trajectory response to the horizontal corrector H101 and vertical corrector V101 before (blue line) and after (magenta line) fit of model parameters compared to experimental measurements (black dots)}
    \label{f_FAST_sampleCorr}
\end{figure}

\begin{figure}[htb]
    \centering
	\includegraphics[width=\columnwidth]{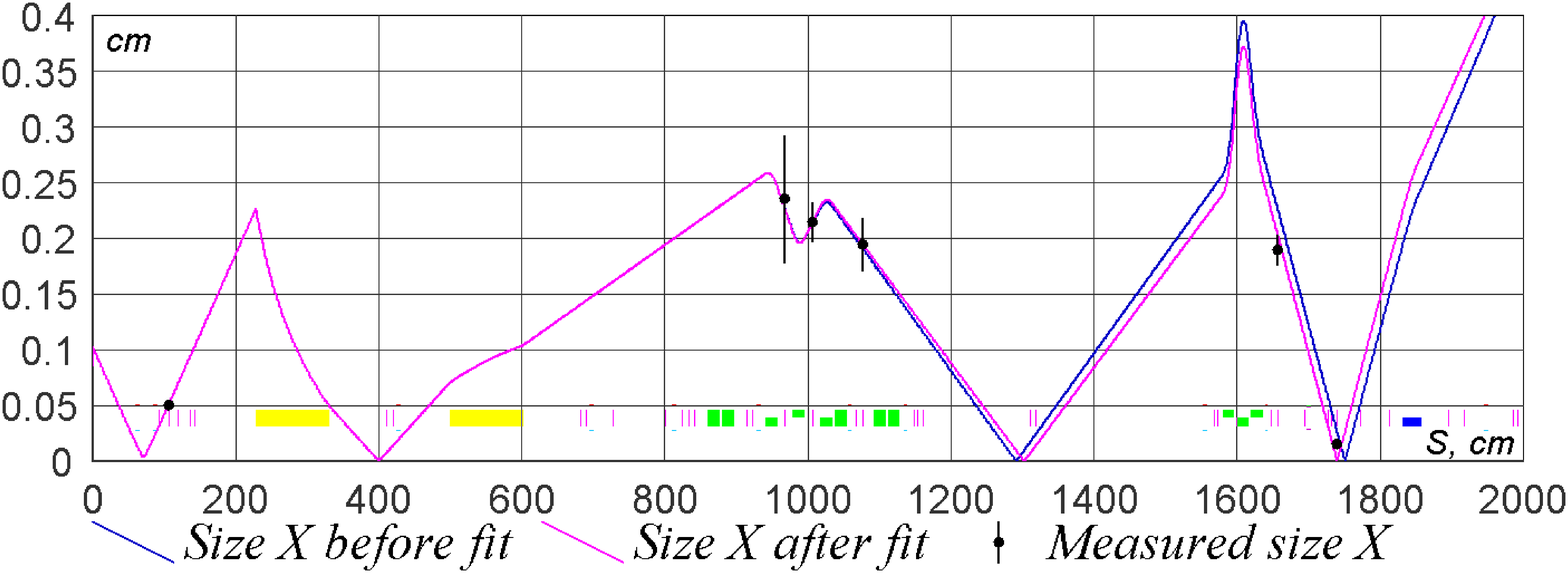}
	\includegraphics[width=\columnwidth]{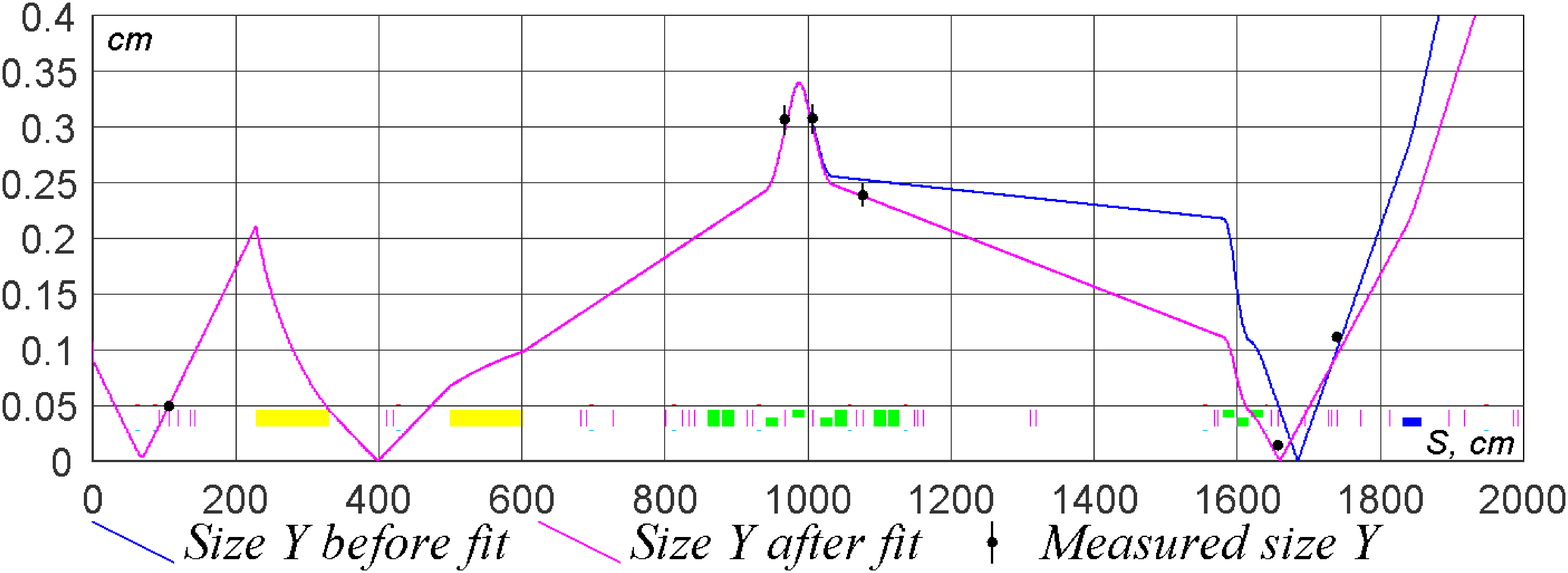}
    \caption{Horizontal and vertical beam envelopes along FAST before (blue line) and after (magenta line) model fit compared to the experimental measurements (black dots)}
    \label{f_FAST_sizes}
\end{figure}

Figure \ref{f_FAST_sampleCorr} shows the difference of the trajectory responses calculated with the initial and the fitted models along with measured data on an example of the vertical and horizontal dipole correctors H101 and V101. Figure \ref{f_FAST_sizes} illustrates the difference in beam envelopes calculated using the model with quadrupole's gradients derived from set currents and from the fitted model for the optimized initial conditions.

\section{Conclusion}

Whenever possible, the design of new machines should incorporate proper distribution of diagnostics and correction elements, and have a provision for adequate correction methods. When such measurement and correction schemes cannot be implemented, for example due to the lack of physical space, the required level of magnetic optics precision can be attained by other means. The enhancement of experimental data can help in such situations. In particular, the proposed measurement of responses of orbit bumps to variation of focusing strengths uses the same instrumentation as that needed for standard LOCO.

All of the discussed algorithms have been implemented in the "Sixdsimulation" software, which is available upon request. "Sixdsimulation" computes fully coupled linear lattice parameters in 6-dimensional phase space and can be integrated into existing accelerator control systems, thus allowing to measure or load experimental data, process it and implement corrections from one program. Flexibility of "Sixdsimulation" is demonstrated by application to several different accelerators. Simulations confirms that the extended LOCO algorithm implemented in the code can be effective in resolving degeneracies, which can save time during commissioning  and increase reliability of existing machines.

\bibliography{latCorr}%

\begin{thebibliography}{15}%
\makeatletter
\providecommand \@ifxundefined [1]{%
 \@ifx{#1\undefined}
}%
\providecommand \@ifnum [1]{%
 \ifnum #1\expandafter \@firstoftwo
 \else \expandafter \@secondoftwo
 \fi
}%
\providecommand \@ifx [1]{%
 \ifx #1\expandafter \@firstoftwo
 \else \expandafter \@secondoftwo
 \fi
}%
\providecommand \natexlab [1]{#1}%
\providecommand \enquote  [1]{``#1''}%
\providecommand \bibnamefont  [1]{#1}%
\providecommand \bibfnamefont [1]{#1}%
\providecommand \citenamefont [1]{#1}%
\providecommand \href@noop [0]{\@secondoftwo}%
\providecommand \href [0]{\begingroup \@sanitize@url \@href}%
\providecommand \@href[1]{\@@startlink{#1}\@@href}%
\providecommand \@@href[1]{\endgroup#1\@@endlink}%
\providecommand \@sanitize@url [0]{\catcode `\\12\catcode `\$12\catcode
  `\&12\catcode `\#12\catcode `\^12\catcode `\_12\catcode `\%12\relax}%
\providecommand \@@startlink[1]{}%
\providecommand \@@endlink[0]{}%
\providecommand \url  [0]{\begingroup\@sanitize@url \@url }%
\providecommand \@url [1]{\endgroup\@href {#1}{\urlprefix }}%
\providecommand \urlprefix  [0]{URL }%
\providecommand \Eprint [0]{\href }%
\providecommand \doibase [0]{http://dx.doi.org/}%
\providecommand \selectlanguage [0]{\@gobble}%
\providecommand \bibinfo  [0]{\@secondoftwo}%
\providecommand \bibfield  [0]{\@secondoftwo}%
\providecommand \translation [1]{[#1]}%
\providecommand \BibitemOpen [0]{}%
\providecommand \bibitemStop [0]{}%
\providecommand \bibitemNoStop [0]{.\EOS\space}%
\providecommand \EOS [0]{\spacefactor3000\relax}%
\providecommand \BibitemShut  [1]{\csname bibitem#1\endcsname}%
\let\auto@bib@innerbib\@empty
\bibitem [{\citenamefont {Safranek}(1997)}]{Safranek:1997mra}%
  \BibitemOpen
  \bibfield  {author} {\bibinfo {author} {\bibfnamefont {J.}~\bibnamefont
  {Safranek}},\ }\bibfield  {title} {\enquote {\bibinfo {title} {{Experimental
  determination of storage ring optics using orbit response measurements}},}\
  }\href {\doibase 10.1016/S0168-9002(97)00309-4} {\bibfield  {journal}
  {\bibinfo  {journal} {Nucl. Instrum. Meth.}\ }\textbf {\bibinfo {volume}
  {A388}},\ \bibinfo {pages} {27--36} (\bibinfo {year} {1997})}\BibitemShut
  {NoStop}%
\bibitem [{\citenamefont {Sajaev}\ \emph {et~al.}(2005)\citenamefont {Sajaev},
  \citenamefont {Lebedev}, \citenamefont {Nagaslaev},\ and\ \citenamefont
  {Valishev}}]{Sajaev:2005qw}%
  \BibitemOpen
  \bibfield  {author} {\bibinfo {author} {\bibfnamefont {V.}~\bibnamefont
  {Sajaev}}, \bibinfo {author} {\bibfnamefont {V.}~\bibnamefont {Lebedev}},
  \bibinfo {author} {\bibfnamefont {V.}~\bibnamefont {Nagaslaev}}, \ and\
  \bibinfo {author} {\bibfnamefont {A.}~\bibnamefont {Valishev}},\ }\bibfield
  {title} {\enquote {\bibinfo {title} {{Fully coupled analysis of orbit
  response matrices at the FNAL Tevatron}},}\ }\bibfield  {booktitle} {\emph
  {\bibinfo {booktitle} {{Particle accelerator. Proceedings, Conference,
  PAC'05, Knoxville, USA, May 16-20, 2005}}},\ }\href@noop {} {\bibfield
  {journal} {\bibinfo  {journal} {Conf. Proc.}\ }\textbf {\bibinfo {volume}
  {C0505161}},\ \bibinfo {pages} {3662} (\bibinfo {year} {2005})},\ \bibinfo
  {note} {[,3662(2005)]}\BibitemShut {NoStop}%
\bibitem [{\citenamefont {Resende}\ \emph {et~al.}(2010)\citenamefont
  {Resende}, \citenamefont {Farias}, \citenamefont {Liu}, \citenamefont
  {Plotegher},\ and\ \citenamefont {Tavares}}]{Resende:2010cfa}%
  \BibitemOpen
  \bibfield  {author} {\bibinfo {author} {\bibfnamefont {Ximenes}\ \bibnamefont
  {Resende}}, \bibinfo {author} {\bibfnamefont {Ruy}\ \bibnamefont {Farias}},
  \bibinfo {author} {\bibfnamefont {Lin}\ \bibnamefont {Liu}}, \bibinfo
  {author} {\bibfnamefont {Matheus}\ \bibnamefont {Plotegher}}, \ and\ \bibinfo
  {author} {\bibfnamefont {Pedro}\ \bibnamefont {Tavares}},\ }\bibfield
  {title} {\enquote {\bibinfo {title} {{Analysis of the LNLS Storage Ring
  Optics Using LOCO}},}\ }in\ \href
  {http://accelconf.web.cern.ch/AccelConf/PAC2009/papers/th6pfp012.pdf} {\emph
  {\bibinfo {booktitle} {Particle accelerator. Proceedings, 23rd Conference,
  PAC'09, Vancouver, Canada, May 4-8, 2009}}}\ (\bibinfo {year} {2010})\ p.\
  \bibinfo {pages} {TH6PFP012}\BibitemShut {NoStop}%
\bibitem [{\citenamefont {Liu}(2010)}]{Liu:2010ula}%
  \BibitemOpen
  \bibfield  {author} {\bibinfo {author} {\bibfnamefont {Guimin}\ \bibnamefont
  {Liu}},\ }\bibfield  {title} {\enquote {\bibinfo {title} {{Linear Optics
  Calibrations for the SSRF Storage Ring Based on COD}},}\ }in\ \href
  {http://accelconf.web.cern.ch/AccelConf/PAC2009/papers/fr5pfp028.pdf} {\emph
  {\bibinfo {booktitle} {{Particle accelerator. Proceedings, 23rd Conference,
  PAC'09, Vancouver, Canada, May 4-8, 2009}}}}\ (\bibinfo {year} {2010})\ p.\
  \bibinfo {pages} {FR5PFP028}\BibitemShut {NoStop}%
\bibitem [{\citenamefont {Benedetti}\ \emph {et~al.}(2011)\citenamefont
  {Benedetti}, \citenamefont {Einfeld}, \citenamefont {Mart?},\ and\
  \citenamefont {Munoz}}]{Benedetti:2011za}%
  \BibitemOpen
  \bibfield  {author} {\bibinfo {author} {\bibfnamefont {G.}~\bibnamefont
  {Benedetti}}, \bibinfo {author} {\bibfnamefont {D.}~\bibnamefont {Einfeld}},
  \bibinfo {author} {\bibfnamefont {Z.}~\bibnamefont {Mart?}}, \ and\ \bibinfo
  {author} {\bibfnamefont {M.}~\bibnamefont {Munoz}},\ }\bibfield  {title}
  {\enquote {\bibinfo {title} {{LOCO in the ALBA Storage Ring}},}\ }\bibfield
  {booktitle} {\emph {\bibinfo {booktitle} {{Particle accelerator. Proceedings,
  2nd International Conference, IPAC 2011, San Sebastian, Spain, September 4-9,
  2011}}},\ }\href@noop {} {\bibfield  {journal} {\bibinfo  {journal} {Conf.
  Proc.}\ }\textbf {\bibinfo {volume} {C110904}},\ \bibinfo {pages}
  {2055--2057} (\bibinfo {year} {2011})}\BibitemShut {NoStop}%
\bibitem [{\citenamefont {Roblin}(2011)}]{Roblin:2011zz}%
  \BibitemOpen
  \bibfield  {author} {\bibinfo {author} {\bibfnamefont {Y.}~\bibnamefont
  {Roblin}},\ }\bibfield  {title} {\enquote {\bibinfo {title} {{Calibrating
  Transport Lines using LOCO Techniques}},}\ }\bibfield  {booktitle} {\emph
  {\bibinfo {booktitle} {{Particle accelerator. Proceedings, 2nd International
  Conference, IPAC 2011, San Sebastian, Spain, September 4-9, 2011}}},\
  }\href@noop {} {\bibfield  {journal} {\bibinfo  {journal} {Conf. Proc.}\
  }\textbf {\bibinfo {volume} {C110904}},\ \bibinfo {pages} {2118--2120}
  (\bibinfo {year} {2011})}\BibitemShut {NoStop}%
\bibitem [{\citenamefont {Aiba}\ \emph {et~al.}(2011)\citenamefont {Aiba},
  \citenamefont {Boge}, \citenamefont {Chrin}, \citenamefont {Milas},
  \citenamefont {Schilcher},\ and\ \citenamefont {Streun}}]{Aiba:2011zb}%
  \BibitemOpen
  \bibfield  {author} {\bibinfo {author} {\bibfnamefont {M.}~\bibnamefont
  {Aiba}}, \bibinfo {author} {\bibfnamefont {M.}~\bibnamefont {Boge}}, \bibinfo
  {author} {\bibfnamefont {J.~T.~M.}\ \bibnamefont {Chrin}}, \bibinfo {author}
  {\bibfnamefont {N.}~\bibnamefont {Milas}}, \bibinfo {author} {\bibfnamefont
  {T.}~\bibnamefont {Schilcher}}, \ and\ \bibinfo {author} {\bibfnamefont
  {A.}~\bibnamefont {Streun}},\ }\bibfield  {title} {\enquote {\bibinfo {title}
  {{Comparison of Linear Optics Correction Means at the SLS}},}\ }\bibfield
  {booktitle} {\emph {\bibinfo {booktitle} {{Particle accelerator. Proceedings,
  2nd International Conference, IPAC 2011, San Sebastian, Spain, September 4-9,
  2011}}},\ }\href@noop {} {\bibfield  {journal} {\bibinfo  {journal} {Conf.
  Proc.}\ }\textbf {\bibinfo {volume} {C110904}},\ \bibinfo {pages}
  {3034--3036} (\bibinfo {year} {2011})}\BibitemShut {NoStop}%
\bibitem [{\citenamefont {Martin}\ \emph {et~al.}(2014)\citenamefont {Martin},
  \citenamefont {Abbott}, \citenamefont {Bartolini}, \citenamefont {Furseman},\
  and\ \citenamefont {Rehm}}]{Martin:2014zpa}%
  \BibitemOpen
  \bibfield  {author} {\bibinfo {author} {\bibfnamefont {Ian}\ \bibnamefont
  {Martin}}, \bibinfo {author} {\bibfnamefont {Michael}\ \bibnamefont
  {Abbott}}, \bibinfo {author} {\bibfnamefont {Riccardo}\ \bibnamefont
  {Bartolini}}, \bibinfo {author} {\bibfnamefont {Matthew}\ \bibnamefont
  {Furseman}}, \ and\ \bibinfo {author} {\bibfnamefont {Guenther}\ \bibnamefont
  {Rehm}},\ }\bibfield  {title} {\enquote {\bibinfo {title} {{A Fast Optics
  Correction for the Diamond Storage Ring}},}\ }in\ \href
  {http://jacow.org/IPAC2014/papers/tupri083.pdf} {\emph {\bibinfo {booktitle}
  {{Proceedings, 5th International Particle Accelerator Conference (IPAC 2014):
  Dresden, Germany, June 15-20, 2014}}}}\ (\bibinfo {year} {2014})\ p.\
  \bibinfo {pages} {TUPRI083}\BibitemShut {NoStop}%
\bibitem [{\citenamefont {Smaluk}\ \emph {et~al.}(2016)\citenamefont {Smaluk},
  \citenamefont {Guo}, \citenamefont {Hidaka}, \citenamefont {Li},
  \citenamefont {Wang}, \citenamefont {Yang},\ and\ \citenamefont
  {Yang}}]{Smaluk:2016vgy}%
  \BibitemOpen
  \bibfield  {author} {\bibinfo {author} {\bibfnamefont {Victor}\ \bibnamefont
  {Smaluk}}, \bibinfo {author} {\bibfnamefont {Weiming}\ \bibnamefont {Guo}},
  \bibinfo {author} {\bibfnamefont {Yoshiteru}\ \bibnamefont {Hidaka}},
  \bibinfo {author} {\bibfnamefont {Yongjun}\ \bibnamefont {Li}}, \bibinfo
  {author} {\bibfnamefont {Guimei}\ \bibnamefont {Wang}}, \bibinfo {author}
  {\bibfnamefont {Lingyun}\ \bibnamefont {Yang}}, \ and\ \bibinfo {author}
  {\bibfnamefont {Xi}~\bibnamefont {Yang}},\ }\bibfield  {title} {\enquote
  {\bibinfo {title} {{Experimental Crosscheck of Algorithms for Magnet Lattice
  Correction}},}\ }in\ \href {\doibase 10.18429/JACoW-IPAC2016-THPMR008} {\emph
  {\bibinfo {booktitle} {{Proceedings, 7th International Particle Accelerator
  Conference (IPAC 2016): Busan, Korea, May 8-13, 2016}}}}\ (\bibinfo {year}
  {2016})\ p.\ \bibinfo {pages} {THPMR008}\BibitemShut {NoStop}%
\bibitem [{\citenamefont {Ji}\ \emph {et~al.}(2016)\citenamefont {Ji},
  \citenamefont {Bai}, \citenamefont {Dutheil}, \citenamefont {Hinder},
  \citenamefont {Lorentz}, \citenamefont {Simon},\ and\ \citenamefont
  {Weidemann}}]{Ji:2016ayy}%
  \BibitemOpen
  \bibfield  {author} {\bibinfo {author} {\bibfnamefont {Daheng}\ \bibnamefont
  {Ji}}, \bibinfo {author} {\bibfnamefont {Mei}\ \bibnamefont {Bai}}, \bibinfo
  {author} {\bibfnamefont {Yann}\ \bibnamefont {Dutheil}}, \bibinfo {author}
  {\bibfnamefont {Fabian}\ \bibnamefont {Hinder}}, \bibinfo {author}
  {\bibfnamefont {Bernd}\ \bibnamefont {Lorentz}}, \bibinfo {author}
  {\bibfnamefont {Michael}\ \bibnamefont {Simon}}, \ and\ \bibinfo {author}
  {\bibfnamefont {Christian}\ \bibnamefont {Weidemann}},\ }\bibfield  {title}
  {\enquote {\bibinfo {title} {{First Experience of Applying Loco for Optics at
  Cosy}},}\ }in\ \href {\doibase 10.18429/JACoW-IPAC2016-TUPMR026} {\emph
  {\bibinfo {booktitle} {{Proceedings, 7th International Particle Accelerator
  Conference (IPAC 2016): Busan, Korea, May 8-13, 2016}}}}\ (\bibinfo {year}
  {2016})\ p.\ \bibinfo {pages} {TUPMR026}\BibitemShut {NoStop}%
\bibitem [{\citenamefont {Huang}\ \emph {et~al.}(2005)\citenamefont {Huang},
  \citenamefont {Lee}, \citenamefont {Prebys},\ and\ \citenamefont
  {Tomlin}}]{Huang:2005gd}%
  \BibitemOpen
  \bibfield  {author} {\bibinfo {author} {\bibfnamefont {X.~B.}\ \bibnamefont
  {Huang}}, \bibinfo {author} {\bibfnamefont {S.~Y.}\ \bibnamefont {Lee}},
  \bibinfo {author} {\bibfnamefont {E.}~\bibnamefont {Prebys}}, \ and\ \bibinfo
  {author} {\bibfnamefont {R.}~\bibnamefont {Tomlin}},\ }\bibfield  {title}
  {\enquote {\bibinfo {title} {{Application of independent component analysis
  to Fermilab Booster}},}\ }\href {\doibase 10.1103/PhysRevSTAB.8.064001}
  {\bibfield  {journal} {\bibinfo  {journal} {Phys. Rev. ST Accel. Beams}\
  }\textbf {\bibinfo {volume} {8}},\ \bibinfo {pages} {064001} (\bibinfo {year}
  {2005})}\BibitemShut {NoStop}%
\bibitem [{\citenamefont {Shatunov}\ \emph {et~al.}(2000)\citenamefont
  {Shatunov} \emph {et~al.}}]{Shatunov:2000zc}%
  \BibitemOpen
  \bibfield  {author} {\bibinfo {author} {\bibfnamefont {{\relax Yu}.~M.}\
  \bibnamefont {Shatunov}} \emph {et~al.},\ }\bibfield  {title} {\enquote
  {\bibinfo {title} {{Project of a new electron positron collider
  VEPP-2000}},}\ }\bibfield  {booktitle} {\emph {\bibinfo {booktitle}
  {{Particle accelerator. Proceedings, 7th European Conference, EPAC 2000,
  Vienna, Austria, June 26-30, 2000. Vol. 1-3}}},\ }\href@noop {} {\bibfield
  {journal} {\bibinfo  {journal} {Conf. Proc.}\ }\textbf {\bibinfo {volume}
  {C0006262}},\ \bibinfo {pages} {439--441} (\bibinfo {year} {2000})},\
  \bibinfo {note} {[,439(2000)]}\BibitemShut {NoStop}%
\bibitem [{\citenamefont {Danilov}\ \emph {et~al.}(1996)\citenamefont
  {Danilov}, \citenamefont {Ivanov}, \citenamefont {Koop}, \citenamefont
  {Nesterenko}, \citenamefont {Perevedentsev}, \citenamefont {Shatilov},
  \citenamefont {Shatunov},\ and\ \citenamefont {Skrinsky}}]{Danilov:1996jw}%
  \BibitemOpen
  \bibfield  {author} {\bibinfo {author} {\bibfnamefont {V.~V.}\ \bibnamefont
  {Danilov}}, \bibinfo {author} {\bibfnamefont {P.~M.}\ \bibnamefont {Ivanov}},
  \bibinfo {author} {\bibfnamefont {I.~A.}\ \bibnamefont {Koop}}, \bibinfo
  {author} {\bibfnamefont {I.~N.}\ \bibnamefont {Nesterenko}}, \bibinfo
  {author} {\bibfnamefont {E.~A.}\ \bibnamefont {Perevedentsev}}, \bibinfo
  {author} {\bibfnamefont {D.~N.}\ \bibnamefont {Shatilov}}, \bibinfo {author}
  {\bibfnamefont {{\relax Yu}.~M.}\ \bibnamefont {Shatunov}}, \ and\ \bibinfo
  {author} {\bibfnamefont {A.~N.}\ \bibnamefont {Skrinsky}},\ }\bibfield
  {title} {\enquote {\bibinfo {title} {{The concept of round colliding
  beams}},}\ }\bibfield  {booktitle} {\emph {\bibinfo {booktitle} {{Particle
  accelerator. Proceedings, 5th European Conference, EPAC 96, Sitges, Spain,
  June 10-14, 1996. Vol. 1-3}}},\ }\href@noop {} {\bibfield  {journal}
  {\bibinfo  {journal} {Conf. Proc.}\ }\textbf {\bibinfo {volume} {C960610}},\
  \bibinfo {pages} {1149--1151} (\bibinfo {year} {1996})},\ \bibinfo {note}
  {[,1149(1996)]}\BibitemShut {NoStop}%
\bibitem [{\citenamefont {Starostenko}\ \emph {et~al.}(2014)\citenamefont
  {Starostenko}, \citenamefont {Levichev},\ and\ \citenamefont
  {Nikiforov}}]{Starostenko:2014iaa}%
  \BibitemOpen
  \bibfield  {author} {\bibinfo {author} {\bibfnamefont {Alexandr}\
  \bibnamefont {Starostenko}}, \bibinfo {author} {\bibfnamefont {Alexey}\
  \bibnamefont {Levichev}}, \ and\ \bibinfo {author} {\bibfnamefont {Danila}\
  \bibnamefont {Nikiforov}},\ }\bibfield  {title} {\enquote {\bibinfo {title}
  {{Commissioning of BINP Injection Complex VEPP-5}},}\ }in\ \href
  {http://jacow.org/LINAC2014/papers/tupp120.pdf} {\emph {\bibinfo {booktitle}
  {{Proceedings, 27th Linear Accelerator Conference, LINAC2014: Geneva,
  Switzerland, August 31-September 5, 2014}}}}\ (\bibinfo {year} {2014})\ p.\
  \bibinfo {pages} {TUPP120}\BibitemShut {NoStop}%
\bibitem [{\citenamefont {Dean~Edstrom}(2016)}]{FAST}%
  \BibitemOpen
  \bibfield  {author} {\bibinfo {author} {\bibfnamefont {et~al.}\ \bibnamefont
  {Dean~Edstrom}},\ }\bibfield  {title} {\enquote {\bibinfo {title} {{50-MeV
  run of IOTA/FAST electron acceleraor}},}\ }in\ \href@noop {} {\emph {\bibinfo
  {booktitle} {{2nd North American Particle Accelerator Conference (NAPAC2016)
  Chicago, Illinois, USA, October 9-14, 2016}}}}\ (\bibinfo {year} {2016})\ p.\
  \bibinfo {pages} {TUPOA19}\BibitemShut {NoStop}%
\end{thebibliography}%

\end{document}